\documentclass[lettersize,journal]{IEEEtran}
\usepackage{amsmath,amsfonts}
\usepackage{array}
\usepackage{subfig}
\usepackage{stfloats}
\usepackage{url}
\usepackage{verbatim}
\usepackage{graphicx}
\usepackage{cite}
\usepackage{microtype}
\usepackage{doi}
\usepackage{float}
\usepackage{mathrsfs}
\usepackage{threeparttable}
\usepackage{booktabs}
\usepackage{multicol}
\usepackage{multirow}
\usepackage{tabto}
\usepackage[ruled]{algorithm2e}
\usepackage{algpseudocode}
\usepackage{makecell}
\usepackage{array}
\usepackage{diagbox}
\usepackage{adjustbox}
\usepackage{hyperref}
\usepackage{textcomp}

\begin{document}
\title{Feasibility Study of a Diffusion-Based Model for Cross-Modal Generation of Knee MRI from X-ray: Integrating Radiographic Feature Information}
\author{Zhe Wang, Yung Hsin Chen, Aladine Chetouani, Fabian Bauer, Yuhua Ru, Fang Chen, Liping Zhang, Rachid Jennane$^*$, Mohamed Jarraya$^*$
\thanks{Zhe Wang, Yung Hsin Chen and Mohamed Jarraya are with Department of Radiology, Massachusetts General Hospital, Harvard Medical School, Boston, 02114, USA (e-mail: zwang78@mgh.harvard.edu; ychen4@mgh.harvard.edu; mjarraya@mgh.harvard.edu).}
\thanks{Aladine Chetouani is with L2TI Laboratory, University Sorbonne Paris Nord, Villetaneuse, 93430, France (e-mail: aladine.chetouani@sorbonne-paris-nord.fr).}
\thanks{Fabian Bauer is with Division of Radiology, German Cancer Research Center, Heidelberg, 69120, Germany (e-mail: fabian.bauer@dkfz-heidelberg.de).}
\thanks{Yuhua Ru is with Jiangsu Institute of Hematology, The First Affiliated Hospital of Soochow University, Suzhou, 215006, China. (e-mail: ruyuhua@163.com).}
\thanks{Fang Chen is with department of Medical School, Henan University of Chinese Medicine, Zhengzhou, 450046, China. (e-mail: chenfangyxy@hactcm.edu.cn).}
\thanks{Liping Zhang is with Athinoula A. Martinos Centre for Biomedical Imaging, Massachusetts General Hospital, Harvard Medical School, Boston, 02114, USA. (e-mail: lzhang90@mgh.harvard.edu).}
\thanks{Rachid Jennane is with IDP Institute, UMR CNRS 7013, University of Orleans, Orleans, 45067, France (e-mail: rachid.jennane@univ-orleans.fr).}}

\markboth{Journal of \LaTeX\ Class Files,~Vol.~14, No.~8, August~2021}%
{Shell \MakeLowercase{\textit{et al.}}: A Sample Article Using IEEEtran.cls for IEEE Journals}

\maketitle
\begin{abstract}
Knee osteoarthritis (KOA) is a prevalent musculoskeletal disorder, often diagnosed using X-rays due to its cost-effectiveness. While Magnetic Resonance Imaging (MRI) provides superior soft tissue visualization and serves as a valuable supplementary diagnostic tool, its high cost and limited accessibility significantly restrict its widespread use. To explore the feasibility of bridging this imaging gap, we conducted a feasibility study leveraging a diffusion-based model that uses an X-ray image as conditional input, alongside target depth and additional patient-specific feature information, to generate corresponding MRI sequences. Our findings demonstrate that the MRI volumes generated by our approach is visually closer to real MRI scans. Moreover, increasing inference steps enhances the continuity and smoothness of the synthesized MRI sequences. Through ablation studies, we further validate that integrating supplementary patient-specific information, beyond what X-rays alone can provide, enhances the accuracy and clinical relevance of the generated MRI, which underscores the potential of leveraging external patient-specific information to improve the MRI generation. This study is available at \url{https://zwang78.github.io/}.
\end{abstract}


\section{Introduction}
\IEEEPARstart{K}{nee} OsteoArthritis (KOA) is a progressive disease marked by the degeneration and damage of articular cartilage, changes at the joint margins, and reactive hyperplasia of the subchondral bone \cite{kneeoa}. Various factors, such as age, obesity, stress, and trauma, can contribute to its development \cite{multi-factor}. Those affected often endure severe pain and limited mobility, which can greatly diminish their quality of life and elevate the risk of chronic conditions like cardiovascular disease \cite{cardiovascular}. Despite extensive research, the precise cause of KOA remains unknown, and there is currently no cure \cite{notclear}.

Traditionally, X-ray imaging has been the most common method for diagnosing KOA due to its accessibility and ability to reveal changes in bone structure, such as Joint Space Narrowing (JSN) and osteophyte formation \cite{sukerkar2022imaging}. To assess KOA on X-rays, five distinct grades (i.e., from G-0 to G-4), determined by the presence and severity of symptoms, are commonly used \cite{KL}. However, X-ray imaging has limitations, particularly in visualizing soft tissue structures and early cartilage changes. Magnetic Resonance Imaging (MRI), on the other hand, provides a more comprehensive assessment of the knee joint by offering detailed images of both bone and soft tissues, including cartilage, menisci, and synovium \cite{eckstein2006magnetic}. MRI can detect early signs of KOA that are not visible on X-rays, such as bone marrow lesions, synovitis, and early cartilage degeneration \cite{piccolo2023imaging}. However, the widespread adoption of MRI in the clinical management of KOA is hindered by several challenges, most notably the high cost associated with MRI scans \cite{reyes2023cost}. The expense of MRI technology, including the costs of the equipment, maintenance, and specialized personnel, makes it less accessible compared to traditional X-ray imaging \cite{bell1996economics}. This economic barrier is particularly significant in resource-limited settings and for patients without adequate health insurance coverage.  In addition to the high financial cost, MRI scans require more time and patient cooperation, as they necessitate the patient to remain still for extended periods \cite{mclean2023mri}, which can be difficult for those experiencing severe pain or limited mobility due to KOA.

To address these limitations, researchers have explored various techniques to enhance MRI reconstruction and reduce its dependency on traditional methods. Fessler et al. \cite{fessler2010model} provides an in-depth review of iterative algorithms used for improving MRI image reconstruction, particularly in scenarios where traditional inverse Fourier transforms are inadequate. The author highlights the limitations of conventional methods, especially when dealing with non-Cartesian sampling, under-sampled data, or the presence of nonlinear magnetic fields, and then focuses on model-based approaches, which utilize iterative methods that incorporate physical models of MRI acquisition, allowing for better image quality by addressing inhomogeneity and under-sampling issues. The authors demonstrate that model-based methods can significantly enhance image reconstruction, providing superior results compared to traditional techniques, though often at the cost of increased computation. In \cite{han2019k}, Han et al. proposed leveraging the k-space to improve the efficiency and accuracy of MRI reconstruction. The authors introduced a framework inspired by the low-rank Hankel matrix completion approach, which directly interpolates missing k-space data and enhances the representation of MRI data by learning the underlying signal structure, resulting in high-quality images even from undersampled data. The proposed k-space deep learning approach significantly reduces the computational complexity and memory requirements. In \cite{lustig2007sparse}, Lustig et al. leveraged compressed sensing to significantly accelerate MRI scan times without compromising image quality. The authors explore the inherent sparsity of MR images, both in the image domain and in the transform domains. By randomly undersampling k-space and using nonlinear recovery schemes, the study shows that high-quality MR images can be reconstructed from incomplete data. This approach effectively reduces acquisition time and mitigates artefacts that arise from undersampling, which demonstrates that compressed sensing has the potential to substantially reduce scan times in clinical settings, making MRI more efficient and accessible.

With the rapid advancements in deep learning technologies, the field of MRI reconstruction has seen significant growth \cite{yang2017dagan} \cite{bahrami2016reconstruction}. Recently, combining multi-modal and diffusion-based models has become a new trend for image generation. In \cite{zhan2024medm2g}, Zhan et al. presented a novel medical generative model, MedM2G, designed to unify multiple medical modalities (e.g., CT, MRI, X-ray) and tasks (e.g., text-to-image, image-to-text, modality translation) into a single, efficient framework. The authors address the challenge of aligning and generating medical multi-modal data, particularly in the absence of large, well-paired datasets. They propose a central alignment strategy to efficiently align different modalities using text as the central modality and introduce medical visual invariant preservation to retain key clinical knowledge from each modality. The model leverages a multi-flow cross-guided diffusion process that facilitates adaptive interaction among modalities, enhancing the generation of accurate and high-quality medical images. In \cite{cola}, Jiang et al. introduced a novel diffusion-based approach to multi-modal MRI synthesis. Traditional diffusion models for MRI synthesis typically operate in the original image domain, leading to high memory consumption and less effective handling of anatomical structures. CoLa-Diff addresses these challenges by operating in the latent space to reduce memory overhead while preserving anatomical details. The model introduces structural guidance using brain region masks and an auto-weight adaptation mechanism to balance multiple input modalities effectively during synthesis. The authors demonstrate that CoLa-Diff outperforms existing state-of-the-art models in multi-modal MRI synthesis tasks, showing superior results in terms of Peak Signal-to-Noise Ratio (PSNR) and Structural Similarity Index (SSIM). In \cite{kim2024adaptive}, Kim et al. propose an advanced method for translating MRI images between different modalities. The authors introduce an Adaptive Latent Diffusion Model (ALDM) that leverages a novel multiple-switchable spatially adaptive normalization (MS-SPADE) block to transform source latent representations into target-like latent. This approach enables high-quality image synthesis across multiple modalities without requiring patch cropping, preserving the global information of the images. By conditioning the model with target modalities and applying the ALDM in a compressed latent space, the proposed method significantly reduces computational complexity while achieving superior results in both one-to-one and one-to-many image translation tasks.

Inspired by the above work, we propose a novel approach based on the Conditional Latent Diffusion Model (CLDM) \cite{rombach2022high}. Specifically, during the training phase, each real MRI slice is first Gaussian noised and then progressively denoised using the corresponding X-ray as conditional input $y$, target depth $d$ and radiographic features $r$ of the specific patient as the guidance module to reconstruct the input MRI slice. In the inference phase, given a knee X-ray image, a target depth $d$ and the radiographic features $r$, the model progressively denoises from Gaussian noise to generate the corresponding MRI slice at that depth. Then, the target depth is adjusted and the denoising process is repeated iteratively for each depth until all the required slices are synthesized. These individual slices are then stacked together to form a fully generated 3D MRI volume. It is noteworthy that the reconstructed MRI is not intended to serve as a clinical replacement for a real MRI. Nevertheless, it signifies a groundbreaking feasibility study in integrating patient-specific radiographic information with knee X-rays to generate corresponding MRI volumes.

The primary contributions of this study are summarized as follows:  
\begin{itemize}  
\item[$\bullet$] A novel CLDM-based framework is proposed for generating MRI volumes from X-ray images.  
\item[$\bullet$] Beyond the target depth, patient-specific radiographic features not present in the X-ray are integrated as guidance to improve the denoising process.
\item[$\bullet$] Ablation studies demonstrate that incorporating additional patient-specific information enhances the accuracy and clinical relevance of the generated MRI.
\item[$\bullet$] By increasing the number of inference steps (i.e., exceeding the depth of the original MRI volume), the model achieves effective interpolation between MRI slices.
\end{itemize}

\section{Proposed approach}
\label{proposed_approach}

\subsection{Classical conditional latent diffusion model}
\begin{figure}[htbp]
\centering
\includegraphics[width=0.48\textwidth]{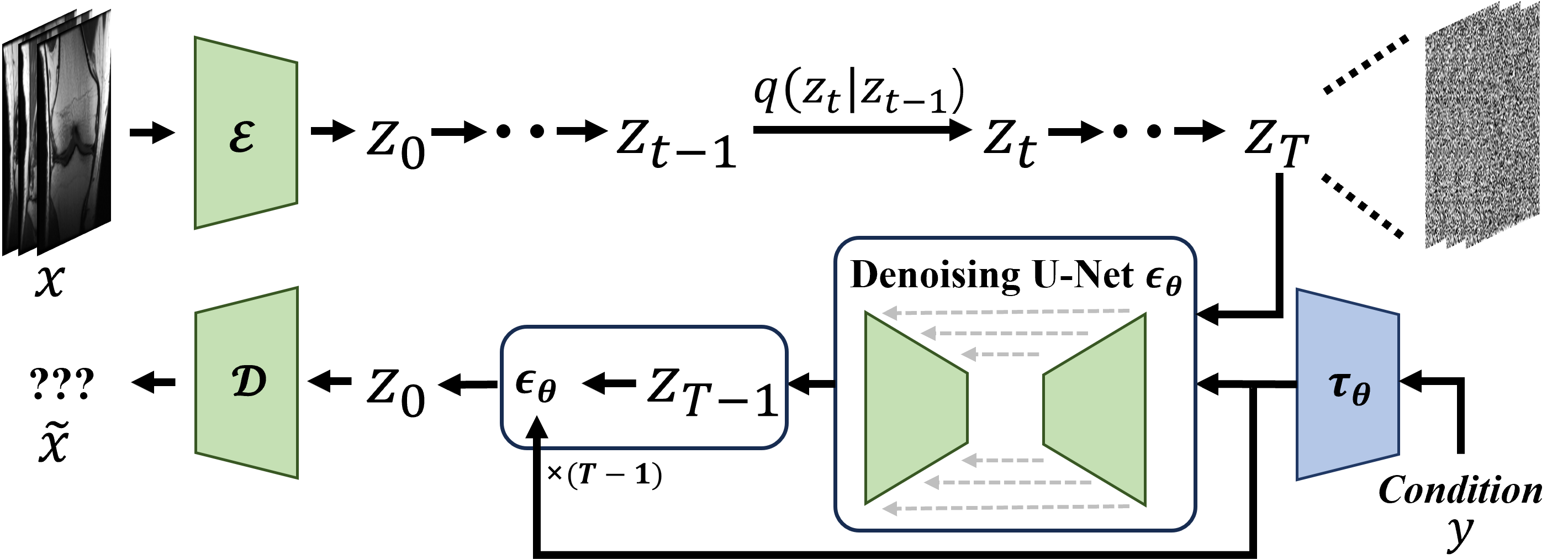}
\caption{The structure of the classical conditional latent diffusion model.}
\label{ddpm}
\end{figure}

Before delving into the proposed approach, we provide a brief overview of the classic conditional latent diffusion model, which serves as the foundation for our methodology. As shown in Fig. \ref{ddpm}, a conditional latent diffusion model is a type of generative model used for producing data (such as images, audio, or text) based on certain conditions or inputs. This model combines the strengths of diffusion models and latent variable models, allowing it to generate complex data distributions while being guided by specific conditions. Different modules are to be introduced, respectively.

\begin{figure*}
\centering 
\includegraphics[width=1\textwidth]{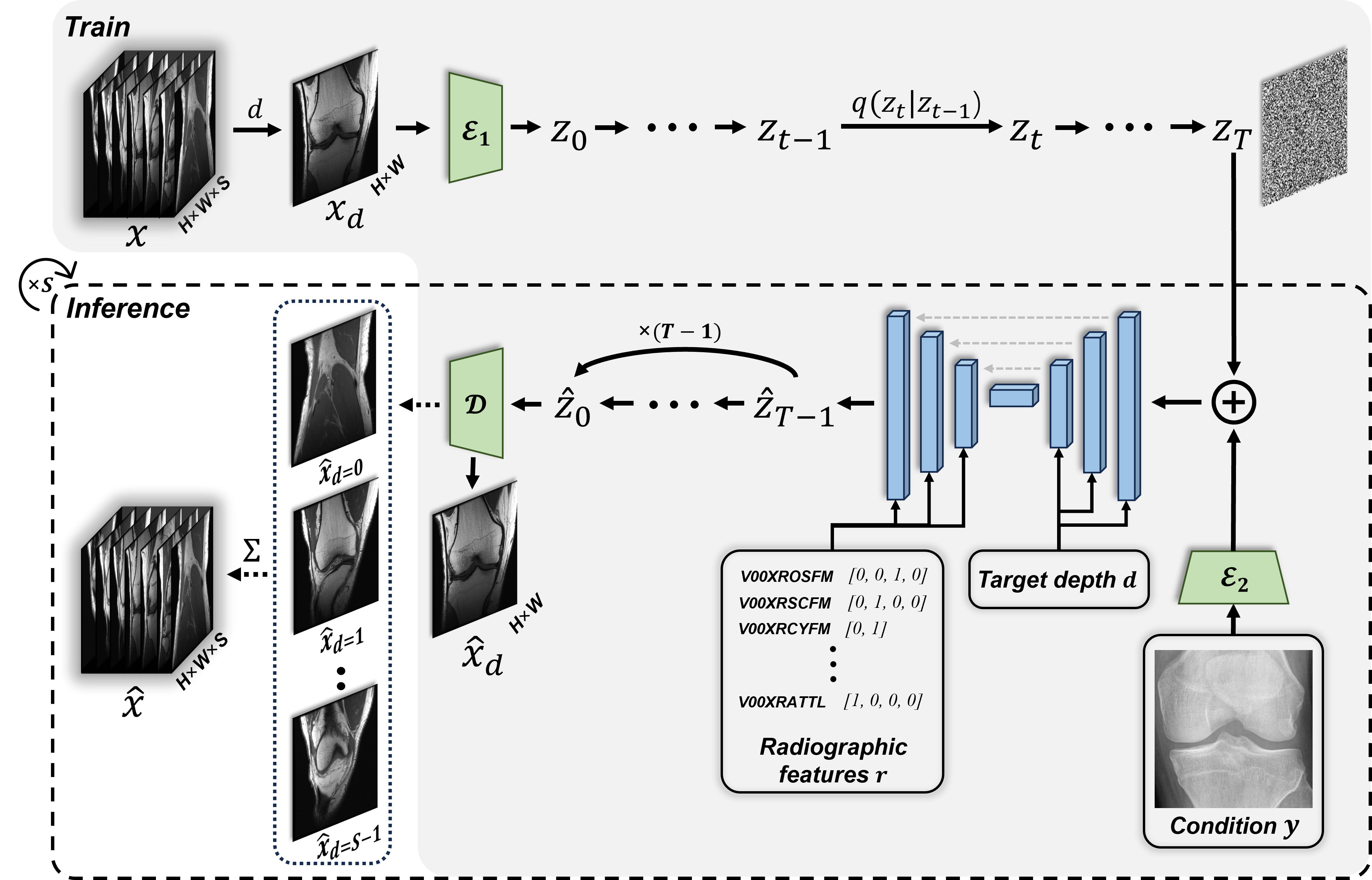}
\caption{The flowchart of the proposed approach, starts by extracting a slice at position $d$ from a real MRI sequence $x$, denoted as $x_d$. This slice is encoded by the encoder $\mathcal{E}_1$ to produce a latent representation $z_0$, which is progressively noised following the distribution $q(z_t | z_{t-1})$, ultimately generating the noised latent representation $z_T$. The corresponding X-ray image $y$ is then used as a conditional input, encoded by $\mathcal{E}_2$, and concatenated with $z_T$ to form the initial input for the downsampling network. During the denoising process, the target depth $d$ and the patient-specific radiographic feature information $r$ are jointly embedded at various stages of the downsampling and upsampling phases, collectively providing denoising guidance. This denoising process is repeated for $T-1$ iterations to progressively refine the latent representation, resulting in the denoised latent $\hat{z}_0$, which is decoded by $\mathcal{D}$ to reconstruct the MRI slice $\hat{x}_d$ at depth $d$. During inference, the process begins with $z_T$ and is repeated $S$ times, corresponding to the number of channels $S$ in the original MRI sequence. Finally, the full generated MRI sequence $\hat{x}$ is obtained by stacking all the generated MRI slices in sequential order (i.e., $\hat{x} = \{\hat{x}_d\}_{d=0}^{S-1}$).}
\label{flowchart}
\end{figure*}

\subsubsection{Encode process}
The latent variable model is used to map the input image $x$ to a latent space using an encoder:

\begin{equation}
z_0 = \mathcal{E}(x)
\end{equation}
where $\mathcal{E}$ is an encoder which maps an image $x$ to a latent representation $z_0$.

\subsubsection{Diffusion process}
A diffusion process describes the evolution of a random variable over time. Through the forward diffusion process, for the original latent representation $z_0 \sim q(z_0)$, each step of the diffusion process, which comprises a total of $T$ steps, involves adding Gaussian noise $n$ to the data obtained from the previous step $z_{t-1}$ as follows:

\begin{equation}
q(z_t\mid z_{t-1})=\mathcal{N} (z_t;\sqrt[]{1-\beta_t}z_{t-1},\beta_t\mathbf{I})
\end{equation}
where $\beta_t$ represents the variance used for each step in the range of $[0,1]$, and $\mathbf{I}$ represents the identity matrix. The entire diffusion process constitutes a Markov chain \cite{norris1998markov}:

\begin{equation}
q(z_{1}\mid z_0)=\prod_{t=1}^{T} q(z_t\mid z_{t-1})
\end{equation}

Here, we directly sample $z_t$ for any step $t$ based on the original data $z_0$: $z_t \sim q(z_t \mid z_0)$ defining $\alpha_t = 1-\beta_t$ and $\bar{\alpha}t = \prod{i=1}^{t} \alpha_i$. Through the reparameterization technique \cite{kingma2013auto}, we have:

\begin{equation}
\begin{aligned}
z_t &= \sqrt{\alpha_t}z_{t-1} + \sqrt{1-\alpha_t}\epsilon_{t-1} \\&= \sqrt  {\alpha _ {t}} (\sqrt  {  \alpha _ {t-1}}  z_ {t-2}  +  \sqrt {1-\alpha _ {t-1}}  \epsilon_{t-2}) +  \sqrt {1-\alpha _ {t}}\epsilon_{t-1}\\&=\sqrt{\bar{\alpha}_t}z_0 + \sqrt{1-\bar{\alpha}_t}\epsilon, \quad \quad \forall \epsilon_t \sim \mathcal{N}(0, \mathbf{I})
\end{aligned}
\label{x_t}
\end{equation}

\subsubsection{Denoising process}
After the diffusion process, denoising is performed by sampling from the updated noise distribution. Estimating distribution $q(x_{t-1}\mid x_t)$ requires the utilization of the entire training set. Typically, a neural network, such as a U-Net \cite{ronneberger2015u}, is employed to estimate these distributions. Here, the reverse process is also defined as a Markov chain composed of a sequence of Gaussian distributions parameterized by neural network parameters:

\begin{equation}
p_\theta(z_{t-1}\mid z_t, y)=\mathcal{N} (z_{t-1};\mu_\theta(z_t,t,y), \sigma_\theta^2(t))
\end{equation}
where $p_\theta(z_{t-1}\mid z_t, y)$ represents parameterized Gaussian distributions, the mean $\mu_\theta(z_t,t,y)$ and variance $\sigma_\theta^2(t)$ at each time step with the condition $y$ are determined by trained learning networks.

\subsubsection{Training process}
The encoder-decoder and diffusion modules are trained separately. For the encoder-decoder, the reconstruction loss $\mathcal{L}_ {rec}$ is as:

\begin{equation}
\label{encoder-decoder-loss}
\mathcal{L}_ {rec} = ||(x, \mathcal{D}(\mathcal{E}(x))||_{2} ^ {2}
\end{equation}

The parameters $\theta$ of the diffusion module aim to minimize the distance between the true noise $\epsilon$ and the predicted noise $\epsilon_\theta$. The diffusion loss $\mathcal{L}_ {diff}$ quantifies this difference as the expected value of the squared norm between the actual noise and the noise estimated by the neural network:

\begin{equation}
\mathcal{L}_ {diff} =  \mathbb{E}_ {z_0 \sim \mathcal{E}(x), y, t, \epsilon \sim \mathcal{N}(0,\mathbf{I})} || \epsilon -  \epsilon_\theta  (z_t,t,y)  ||_{2} ^ {2}
\label{diff_classical_loss}
\end{equation}
where $\mathbb{E}$ represents the expectation over all data points under the joint distribution of the conditioning information $y$, time steps $t$, and Gaussian noise $\epsilon$.

\subsection{Proposed approach}
\label{proposed_model}
As presented in Fig. \ref{flowchart}, our approach consists of three main modules: an auto-encoder module, a CLDM module, and a guidance module. $x_d$ is a slice of the original knee MRI sequence $x$ at the depth position $d$ with the size of $H \times W \times S$. $\mathcal{E}_1$ and $\mathcal{D}$ are encoder and decoder of the AutoencoderKL \cite{kingma2013auto}. Unlike traditional autoencoders that map input data to a deterministic latent space, the AutoencoderKL introduces a probabilistic approach by encoding the input into a Gaussian distribution over the latent space. During training, the Kullback-Leibler (KL) divergence is used as a regularization term to ensure that the learned latent space closely follows a standard normal distribution, promoting smoother and more continuous transitions between latent representations. Specifically, the encoder operates at a resolution of 256$\times$256$\times$3 and begins with a convolutional layer configured with 128 channels and then proceeds through a series of stages, each with 2 residual blocks. The channel dimensions are multiplied according to the sequence 1, 2, 4, and 4. The decoder mirrors the structure of the encoder. The architecture is designed with no dropout, focusing on retaining all features without stochastic regularization during training. Simultaneously, the noised latent code $z_T$ is calculated following Eq. \ref{x_t}:
\begin{equation}
z_T = \sqrt{\bar{\alpha}_t}\mathcal{E}_1({x_d}) + \sqrt{1-\bar{\alpha}_T}\epsilon
\end{equation}

Conditional image $y$ is a corresponding knee joint X-ray image, which is then encoded by the encoder $\mathcal{E}_2$ and concatenated with the $z_T$ as the input of the U-Net, serves as the beginning of the reverse process. The employed 2D U-Net architecture adopts a multi-scale hierarchical structure comprising a time embedding module, an encoder module, a middle module, and a decoder module. This UNet model takes 8-channel input and produces 4-channel output, built upon a base channel configuration of 320. The encoder module begins with a 2D convolutional layer featuring a kernel size of 3 and padding of 1, followed by a series of residual blocks with progressively increasing feature dimensions. Each residual block includes multiple convolutional layers, GroupNorm \cite{wu2018group} for normalization, the Swish activation function \cite{ramachandran2017searching} for non-linearity, and optional dropout for regularization. Two residual blocks are employed per downsampling stage, with attention mechanisms activated at resolutions of 4, 2, and 1. The channel count increases across network layers following a multiplier sequence of 1, 2, 4, and 4. The decoder mirrors this structure, utilizing the same residual blocks to progressively upsample feature maps, reducing the number of channels in reverse order while merging them with corresponding encoder feature maps through skip connections. Additionally, the model incorporates 8 attention heads and employs a spatial transformer with a depth of 1, conditioned on a context dimension of 768.

Next, the target depth $d$ is embedded into the downsampling stage to allow the model to condition the latent representation with spatial hierarchy early in the process, which ensures that depth-related features, such as different tissue layers and bone structures, are incorporated from the outset, preserving the anatomical coherence of the volume. In the upsampling stage, the patient-specific radiographic feature information $r$ (refer to Table \ref{features}) is embedded to guide the synthesis of fine clinical features based on the spatial hierarchy established during downsampling, such as the presence of cysts and the degree of attrition. Specifically, the depth $d$ is normalized to the range $[0, 1]$. On the other hand, the patient-specific radiographic feature information $r$ is encoded using a one-hot encoding scheme.

\begin{table}[htbp]
\centering
\caption{Description of radiographic features}
\label{features}
\setlength{\tabcolsep}{1.3mm} 
\begin{threeparttable}
\begin{tabular}{ll}
\toprule
\textbf{Variable} & \textbf{Description} \\
\midrule
V00XROSFM & Osteophytes (OARSI 0-3), femur medial compartment \\ \midrule
V00XRSCFM & Sclerosis (OARSI 0-3), femur medial compartment \\ \midrule
V00XRCYFM & Cysts (OARSI 0-1), femur medial compartment \\ \midrule
V00XRJSM & Joint space narrowing (OARSI 0-3), medial compartment \\ \midrule
V00XRCHM & Chondrocalcinosis (Grades 0-1), medial compartment \\ \midrule
V00XROSTM & Osteophytes (OARSI 0-3), tibia medial compartment \\ \midrule
V00XRSCTM & Sclerosis (OARSI 0-3), tibia medial compartment\\ \midrule
V00XRCYTM & Cysts (OARSI 0-1), tibia medial compartment \\ \midrule
V00XRATTM & Attrition (OARSI 0-1), tibia medial compartment \\ \midrule
V00XRKL & Kellgren and Lawrence (0-4) for osteoarthritis severity \\ \midrule
V00XROSFL & Osteophytes (OARSI 0-3), femur lateral compartment \\ \midrule
V00XRSCFL & Sclerosis (OARSI 0-3), femur lateral compartment\\ \midrule
V00XRCYFL & Cysts (OARSI 0-1), femur lateral compartment \\ \midrule
V00XRJSL & Joint space narrowing (OARSI 0-3), lateral compartment \\ \midrule
V00XRCHL & Chondrocalcinosis (Grades 0-1), lateral compartment \\ \midrule
V00XROSTL & Osteophytes (OARSI 0-3), tibia lateral compartment \\ \midrule
V00XRSCTL & Sclerosis (OARSI 0-3), tibia lateral compartment \\ \midrule
V00XRCYTL & Cysts (OARSI 0-1), tibia lateral compartment \\ \midrule
V00XRATTL & Attrition (OARSI 0-1), tibia lateral compartment \\
\bottomrule
\end{tabular}
\begin{tablenotes}
\footnotesize
\item[$*$] OARSI: Osteoarthritis Research Society International Grading System
\end{tablenotes}
\end{threeparttable}
\end{table}

During the denoising process, the latent variable $z_{t-1}$ is sampled using the Denoising Diffusion Implicit Model (DDIM) \cite{song2020denoising}, which enables deterministic or stochastic generation depending on the chosen schedule. DDIM refines the sampling process by approximating the reverse diffusion trajectory more efficiently than traditional diffusion models. Specifically, it leverages a non-Markovian process that skips some diffusion steps, allowing for faster inference without compromising the quality of the generated images. The sampling of $z_{t-1}$ is computed as follows:
\begin{equation}
z_{t-1} = \sqrt{\bar{\alpha}_{t-1}} \left(\frac{z_t - \sqrt{1 - \bar{\alpha}_t} \mathbf{\epsilon}_\theta}{\sqrt{\bar{\alpha}_t}}\right) + \sqrt{1 - \bar{\alpha}_{t-1}} \mathbf{\epsilon}_\theta
\end{equation}

As presented before, the diffusion model is implemented through a U-Net $\mathcal{U}$, which is trained based on Eq. \ref{diff_classical_loss}. In this study, the goal function is computed as follows:
\begin{equation}
\label{DDIMM}
\mathop{\min}_{\theta} \mathbb{E}_ {z \sim \varepsilon_1(x_d), t, \epsilon \sim \mathcal{N}(0,\mathbf{I})} || \epsilon -  \epsilon_\theta [z_t,t, (z_T, y, d, r)]  ||_{2} ^ {2}
\end{equation}

Finally, for each reconstructed slice at the depth $d$, denoted as $\hat{x}_d$, it is generated through a decoder $\mathcal{D}$. During the inference phase, this process is repeated $S$ times, corresponding to the depth (i.e., channel) $S$ of the original MRI volume. The final generated MRI sequence $\hat{x}$ is formed by stacking all generated slices, represented as $\hat{x} = \{\hat{x}_d\}_{d=0}^{S-1}$. For more clarity, Algorithm \ref{algorithm1} and Algorithm \ref{algorithm2} describe the training and inference processes of the proposed approach, respectively.

\begin{algorithm}[htbp]
\caption{Training process of the approach}
\textbf{Input:} $x$, $t$, $y$, $r$, $S$,  $\mathcal{E}_2$, and all initial parameters\\
\textbf{Output:}  $\hat{x}_d$, $\mathcal{E}_1$, $\mathcal{D}$, $\mathcal{U}$ and learned parameters\\
\While{\textnormal{\textit{not converged}}}{
\For{$d$ in $\{0,1,...,S-1\}$}{
\textbf{Get} $x_d \leftarrow x[d]$\\
\textbf{Get} $\hat{x}_d \leftarrow \mathcal{D}(\mathcal{E}_1({x_d}))$\\
\textbf{Compute} $\mathcal{L}_{rec}(x_d, \hat{x}_d)$ \Comment{Eq. \ref{encoder-decoder-loss}}\\
\textbf{Update} Parameters of $\mathcal{E}_1$ and $\mathcal{D}$\\}
}
\textbf{Freeze} Parameters of $\mathcal{E}_1$ and $\mathcal{D}$\\
\While{\textnormal{\textit{not converged}}}{
\For{$d$ in $\{0,1,...,S-1\}$}{
\textbf{Get} $x_d \leftarrow x[d]$\\
\textbf{Get} $\epsilon,  z_T \leftarrow noise(\mathcal{E}_1({x_d}), t)$ \Comment{Eq. \ref{x_t}}\\
\textbf{Get} $\epsilon_\theta, \hat{z}_d \leftarrow \mathcal{U}(z_T, \mathcal{E}_2(y), t, d,r)$\\
\textbf{Compute} $\mathcal{L}_{diff}(\epsilon, \epsilon_\theta)$ \Comment{Eq. \ref{DDIMM}}\\
\textbf{Get} $\hat{x}_d \leftarrow \mathcal{D}(\hat{z}_d)$\\
\textbf{Compute} $\mathcal{L}_{rec}(x_d, \hat{x}_d)$ \Comment{Eq. \ref{encoder-decoder-loss}}\\
\textbf{Update} Parameters of $\mathcal{U}$\\}
}
\label{algorithm1}
\end{algorithm}

\begin{algorithm}[htbp]
\caption{Inference process of the approach}
\textbf{Input:}  $t$, $r$, $y$, $S$,  $\mathcal{E}_2$, $\mathcal{D}$ and $\mathcal{U}$\\
\textbf{Output:} $\hat{x}_d$ and $\hat{x}$\\
\textbf{Get} $\epsilon \leftarrow \mathcal{N}(0, \mathbf{I})$\\
\For{$d$ in $\{0,1,...,S-1\}$}{
\textbf{Get} $\hat{z}_d \leftarrow \mathcal{U}(\epsilon, \mathcal{E}_2(y), t, d,r)$\\
\textbf{Get} $\hat{x}_d \leftarrow \mathcal{D}(\hat{z}_d)$\\}
\textbf{Get} $\hat{x} \leftarrow \{\hat{x}_d\}_{d=0}^{S-1}$\\
\label{algorithm2}
\end{algorithm}

\section{Experimental settings}
This section provides an overview of the experimental data, the preprocessing steps undertaken, and the specifics of the experimental procedures.

\subsection{Employed database}
The OsteoArthritis Initiative (OAI) \cite{OAI} represents a significant and rich source of data for researchers investigating KOA and related conditions. By analyzing data from 4,796 participants aged between 45 and 79 years over 96 months, the initiative provides a comprehensive longitudinal dataset. Each participant underwent nine follow-up examinations, allowing for detailed tracking of the disease's progression and the identification of potential risk factors associated with KOA development or progression.


\subsection{Data preprocessing}
\label{data_preprocessing}

\subsubsection{X-ray} 
\begin{figure}[htbp]
\centering
\subfloat[]{
\includegraphics[width=0.2\textwidth]{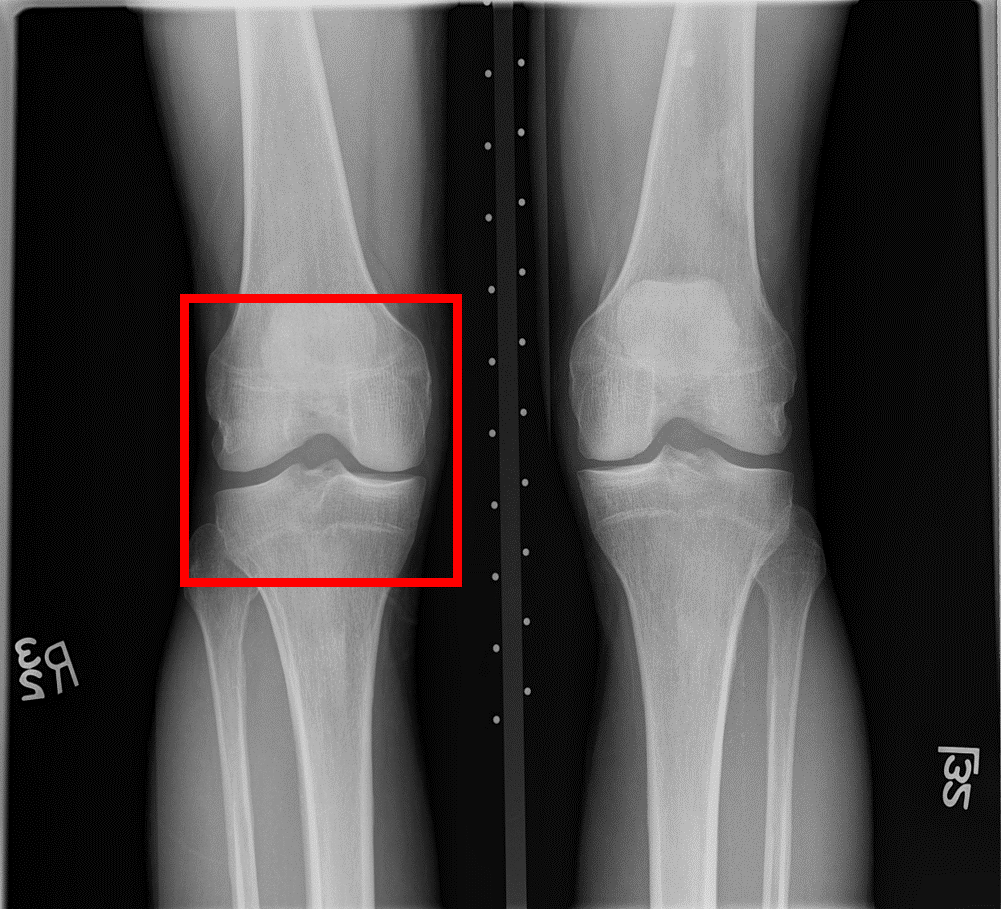}
\label{plainXray}
}
\subfloat[]{
\label{kneeROI}
\begin{minipage}[t]{0.19\textwidth}
\centering
\includegraphics[width=0.96\textwidth]{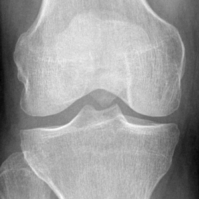}
\end{minipage}
}
\caption{A standard knee plain radiograph and an identified knee joint highlighted in the red box \ref{plainXray}. An identified knee joint \ref{kneeROI}.}
\label{detectionknee}
\end{figure}

As shown in Fig. \ref{detectionknee}, the knee joint (Fig. \ref{kneeROI}) was identified from the plain radiograph (Fig. \ref{plainXray}) through the YOLOv2 learning model \cite{yolov2}, with images sized of 299 $\times$ 299 $\times$ 1, which were resized to the size of 256$\times$256$\times$1, and image intensity distribution was normalized to [-1, 1].

\subsubsection{MRI} 
\begin{figure}[htbp]
\centering
\subfloat[]{
\label{Distribution}
\centering
\includegraphics[width=0.231\textwidth]{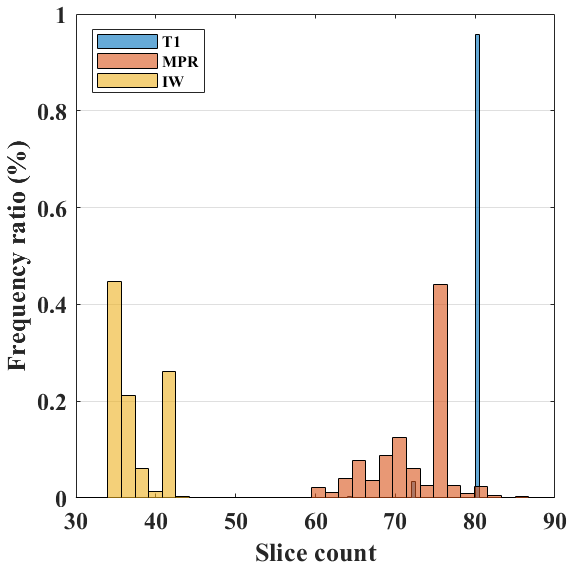}
}
\subfloat[]{
\label{Distribution_}
\centering
\includegraphics[width=0.231\textwidth]{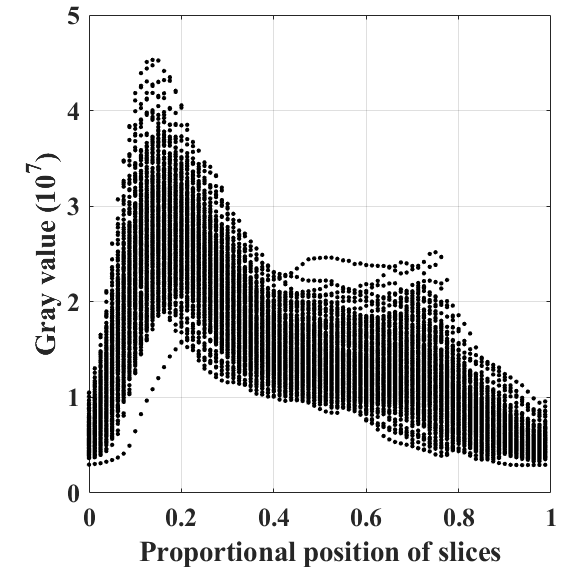}
}
\caption{Distribution of the number of slices for each type of MRI sequence \ref{Distribution}. Grey value distribution at each proportional position of slices within the T1-weighted MRI sequences \ref{Distribution_}.}
\end{figure}

After normalization and standardization, each MRI slice has a size of 256$\times$256$\times$1. To meet the requirements of the pre-trained diffusion model with a channel size of 3, the grayscale images were duplicated three times and then concatenated along the channel dimension. Moreover, the distribution of slice counts was analyzed across three pulse sequences of Intermediate-Weighted (IW), T1-weighted (T1), and Magnetization Prepared Rapid Acquisition Gradient Echo (MPR). As shown in Fig. \ref{Distribution}, the slice count for the T1 sequence remains almost consistently fixed at 80, the MPR sequence displays a broader distribution, ranging from 60 to 80 slices, while the IW sequence has the lowest slice counts, between 30 and 40, which suggests that T1 slices are more uniform and contain richer information for model learning. To ensure consistency in our study, we focused exclusively on T1-weighted MRI scans with 80 slices as the original dataset. Fig. \ref{Distribution_} illustrates the image intensity distribution within each T1 MRI sequence. As can be seen, grey values rise sharply at first, peaking around the 0.15 proportional slice position. After this peak, the grey values drop steeply by the 0.4 proportional slice position and then gradually taper off, eventually returning to levels similar to those of the initial slices. Based on this and subjective assessment, we extracted slices between the 16$\%$ to 78$\%$ proportional positions (i.e., from the 13th slice to the 63rd one) from each MRI sequence to ensure consistency and avoid low information slices at the beginning and end of the sequence.

\subsubsection{Radiographic features} 
The encoding process for the radiographic features involves transforming each feature into a One-Hot vector representation based on its grades. For missing values, all positions in the One-Hot vector are set to zero to indicate the absence of valid information for that feature, which ensures that the model can distinguish missing values from valid inputs without introducing artificial bias. The resulting One-Hot vectors for all radiographic features are concatenated to form a unified representation, which is subsequently used as part of the guidance input during the denoising process in the diffusion model.

After the above processing, the slice count was fixed at 50 for the T1-weighted MRI scans as the final experimental dataset. The resulting 4,262 X-ray-MRI pairs with respective radiographic feature information were then randomly divided into training, validation and testing sets with a ratio of 7:2:1.

\subsection{Experimental details}
The diffusion module utilized pre-trained weights from Zero-123 \cite{Liu_2023_ICCV}. The pre-trained encoder $\mathcal{E}_2$ is sourced from \cite{pmlr-v139}. During the training, a base learning rate of 1e-06 was applied, with a lambda linear scheduler incorporating a warm-up period every 100 steps. The batch size was set to 64, with $T$ configured to 2,000 steps, and an Exponential Moving Average (EMA) \cite{hunter1986exponentially} of 0.99 was used for the diffusion sampling process. The scale factor of the latent space was set as 0.2. The AdamW optimizer \cite{loshchilov2019} was utilized for training over 5,000 epochs. We implemented our approach using PyTorch v1.8.1 \cite{pytorch} on Nvidia A100 graphics cards with 80 GB of memory.

\section{Results and discussion}
\begin{table*}
\centering
\caption{Visualization of a representative generated MRI slice}
\label{vlisualization}
\begin{threeparttable}
\setlength{\tabcolsep}{0.8mm}
\begin{tabular}{rccccc}
\toprule
\multicolumn{1}{c}{X-ray input} & MRI Ground Truth$^*$ &  LDM & Cola-diff & X-diff  & Ours\\
\midrule
\begin{minipage}[b]{0.31\columnwidth}\centering \raisebox{-.5\height}{\includegraphics[width=\linewidth]{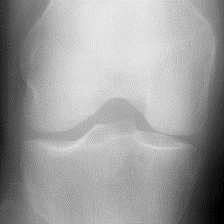}}\end{minipage}&\begin{minipage}[b]{0.31\columnwidth}\centering \raisebox{-.5\height}{\includegraphics[width=\linewidth]{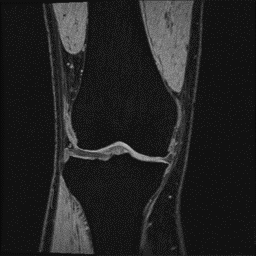}}\end{minipage}&\begin{minipage}[b]{0.31\columnwidth}\centering \raisebox{-.5\height}{\includegraphics[width=\linewidth]{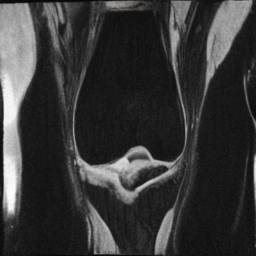}}\end{minipage}&\begin{minipage}[b]{0.31\columnwidth}\centering \raisebox{-.5\height}{\includegraphics[width=\linewidth]{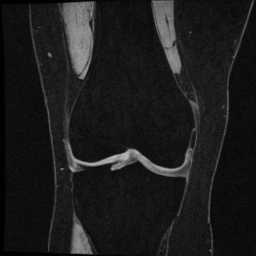}}\end{minipage}&\begin{minipage}[b]{0.31\columnwidth}\centering \raisebox{-.5\height}{\includegraphics[width=\linewidth]{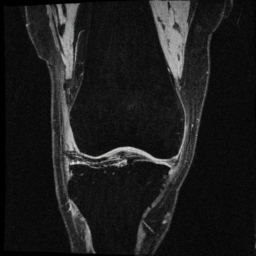}}\end{minipage}&\begin{minipage}[b]{0.31\columnwidth}\centering \raisebox{-.5\height}{\includegraphics[width=\linewidth]{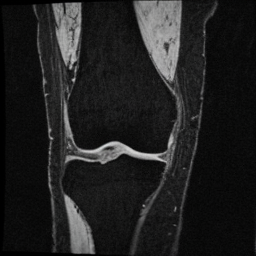}}\end{minipage}\\
\cmidrule(lr){2-6}
\multicolumn{1}{c}{\multirow{1}*[0.49in]{G-0}}\rotatebox[origin=c]{90}{Difference}&\begin{minipage}[b]{0.31\columnwidth}\centering \raisebox{-.5\height}{\includegraphics[width=\linewidth]{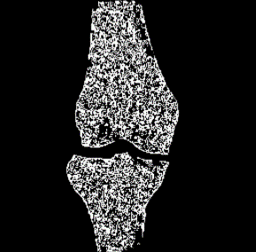}}\end{minipage}&\begin{minipage}[b]{0.31\columnwidth}\centering \raisebox{-.5\height}{\includegraphics[width=\linewidth]{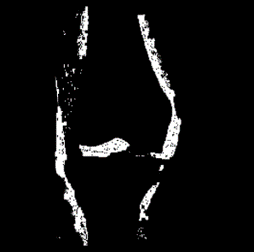}}\end{minipage} &\begin{minipage}[b]{0.31\columnwidth}\centering \raisebox{-.5\height}{\includegraphics[width=\linewidth]{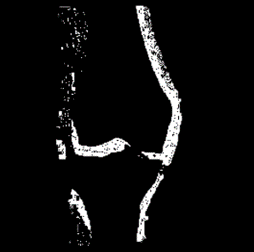}}\end{minipage}&\begin{minipage}[b]{0.31\columnwidth}\centering \raisebox{-.5\height}{\includegraphics[width=\linewidth]{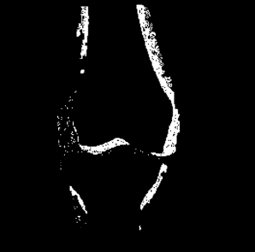}}\end{minipage}&\begin{minipage}[b]{0.31\columnwidth}\centering \raisebox{-.5\height}{\includegraphics[width=\linewidth]{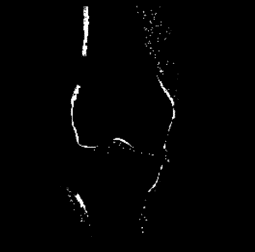}}\end{minipage}\\
\midrule
\begin{minipage}[b]{0.31\columnwidth}\centering \raisebox{-.5\height}{\includegraphics[width=\linewidth]{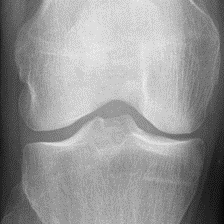}}\end{minipage}&\begin{minipage}[b]{0.31\columnwidth}\centering \raisebox{-.5\height}{\includegraphics[width=\linewidth]{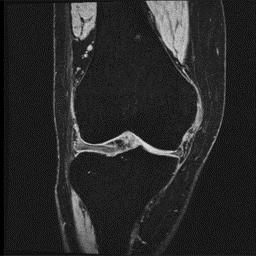}}\end{minipage}&\begin{minipage}[b]{0.31\columnwidth}\centering \raisebox{-.5\height}{\includegraphics[width=\linewidth]{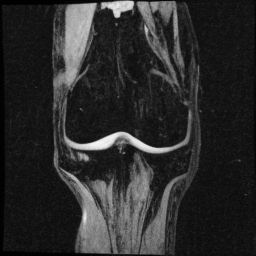}}\end{minipage}&\begin{minipage}[b]{0.31\columnwidth}\centering \raisebox{-.5\height}{\includegraphics[width=\linewidth]{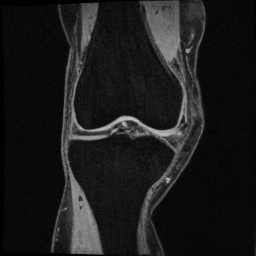}}\end{minipage}&\begin{minipage}[b]{0.31\columnwidth}\centering \raisebox{-.5\height}{\includegraphics[width=\linewidth]{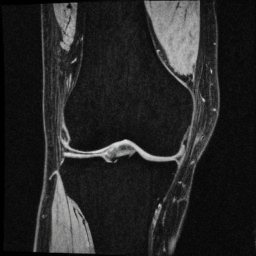}}\end{minipage}&\begin{minipage}[b]{0.31\columnwidth}\centering \raisebox{-.5\height}{\includegraphics[width=\linewidth]{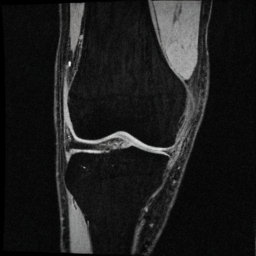}}\end{minipage}\\
\cmidrule(lr){2-6}
\multicolumn{1}{c}{\multirow{1}*[0.47in]{G-2}}\rotatebox[origin=c]{90}{Difference} &\begin{minipage}[b]{0.31\columnwidth}\centering \raisebox{-.5\height}{\includegraphics[width=\linewidth]{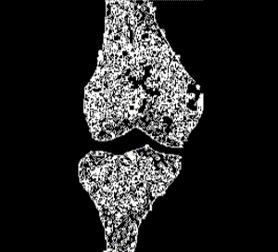}}\end{minipage}&\begin{minipage}[b]{0.31\columnwidth}\centering \raisebox{-.5\height}{\includegraphics[width=\linewidth]{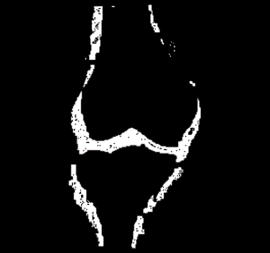}}\end{minipage}&\begin{minipage}[b]{0.31\columnwidth}\centering \raisebox{-.5\height}{\includegraphics[width=\linewidth]{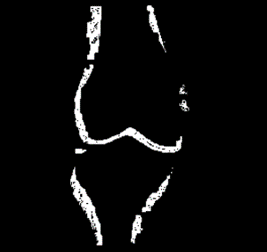}}\end{minipage}&\begin{minipage}[b]{0.31\columnwidth}\centering \raisebox{-.5\height}{\includegraphics[width=\linewidth]{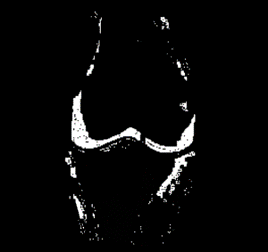}}\end{minipage}&\begin{minipage}[b]{0.31\columnwidth}\centering \raisebox{-.5\height}{\includegraphics[width=\linewidth]{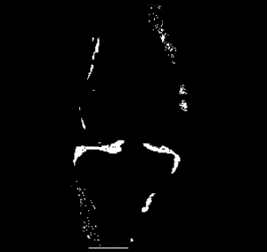}}\end{minipage}\\
\midrule
\begin{minipage}[b]{0.31\columnwidth}\centering \raisebox{-.5\height}{\includegraphics[width=\linewidth]{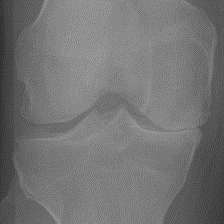}}\end{minipage}&\begin{minipage}[b]{0.31\columnwidth}\centering \raisebox{-.5\height}{\includegraphics[width=\linewidth]{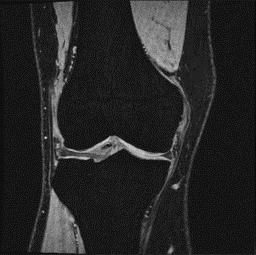}}\end{minipage}&\begin{minipage}[b]{0.31\columnwidth}\centering \raisebox{-.5\height}{\includegraphics[width=\linewidth]{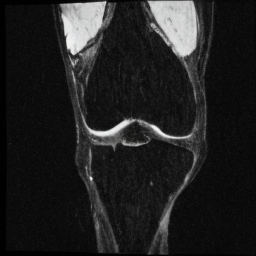}}\end{minipage}&\begin{minipage}[b]{0.31\columnwidth}\centering \raisebox{-.5\height}{\includegraphics[width=\linewidth]{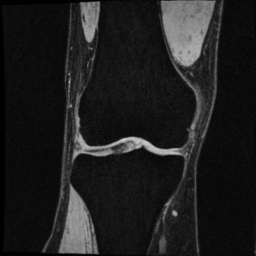}}\end{minipage}&\begin{minipage}[b]{0.31\columnwidth}\centering \raisebox{-.5\height}{\includegraphics[width=\linewidth]{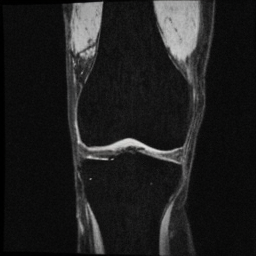}}\end{minipage}&\begin{minipage}[b]{0.31\columnwidth}\centering \raisebox{-.5\height}{\includegraphics[width=\linewidth]{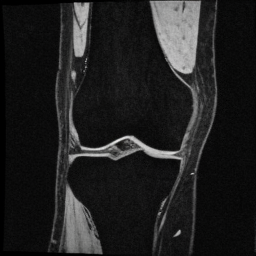}}\end{minipage}\\
\cmidrule(lr){2-6}
\multicolumn{1}{c}{\multirow{1}*[0.47in]{G-4}}\rotatebox[origin=c]{90}{Difference}& \begin{minipage}[b]{0.31\columnwidth}\centering \raisebox{-.5\height}{\includegraphics[width=\linewidth]{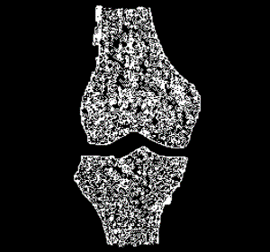}}\end{minipage}&\begin{minipage}[b]{0.31\columnwidth}\centering \raisebox{-.5\height}{\includegraphics[width=\linewidth]{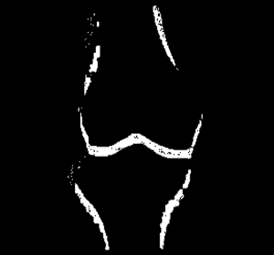}}\end{minipage}&\begin{minipage}[b]{0.31\columnwidth}\centering \raisebox{-.5\height}{\includegraphics[width=\linewidth]{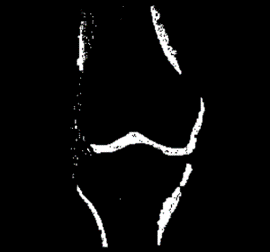}}\end{minipage}&\begin{minipage}[b]{0.31\columnwidth}\centering \raisebox{-.5\height}{\includegraphics[width=\linewidth]{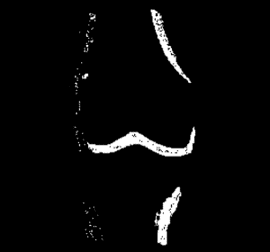}}\end{minipage}&\begin{minipage}[b]{0.31\columnwidth}\centering \raisebox{-.5\height}{\includegraphics[width=\linewidth]{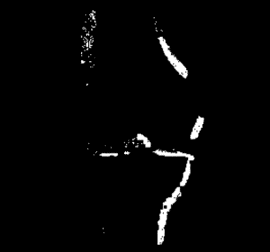}}\end{minipage}\\
\bottomrule
\end{tabular}
\begin{tablenotes}
\footnotesize
\item[$*$] Here, to make it more intuitive and convenient, we only display one slice from each MRI sequence that best represents the complete knee joint.
\end{tablenotes}
\end{threeparttable}
\end{table*}


\subsection{Comparisons with State-of-the-Art Methods}
The comparison is divided into two main parts: Qualitative visualization analysis and quantitative metric-based analysis.

\subsubsection{Qualitative visualization analysis}
Table \ref{vlisualization} provides a comprehensive visual comparison of corresponding generated MRI slices generated by various methods, using X-ray images as input. To facilitate a clear and focused analysis, we selected three representative stages of KOA: G-0, indicating healthy knees; G-2, reflecting early-stage KOA; and G-4, corresponding to late-stage KOA. The compared methods include classic LDM \cite{latent_diffusion_model}, Cola-diff \cite{cola}, X-diff \cite{bourigault2024x}, and our proposed approach. Each row displays the X-ray input on the far left, followed by the real MRI slice and the MRI slices generated by the respective methods. To provide a more intuitive illustration of each method's performance, difference maps generated using a Sobel operator filter are shown beneath each slice, highlighting pixel-wise deviations from the ground truth MRI by detecting edge differences. The visual results consistently show that our model produces generated MRI slices that are closer to the ground truth, with significantly fewer and less pronounced deviations compared to other methods. Specifically, LDM suffers from the lack of constraints provided by the input X-ray. Without this conditional guidance, the model struggles to capture the detailed structural information of the knee joint, leading to great deviations from the ground truth MRI. On the other hand, compared with LDM, although Cola-Diff and X-Diff generate bone structures that are closer to the real MRI in terms of overall shape, they both lack a significant amount of fine details. This limitation prevents them from accurately capturing and reproducing the KOA symptoms. Moreover, it is noteworthy that it is particularly evident as the KOA grade of the input X-ray increases. The progressive narrowing of the knee joint space and the formation of osteophytes are learned and generated more accurately by our model, which demonstrates that our method not only ensures the generation of high-quality MRI but also maintains clinical relevance. However, the model's predictions in areas outside the skeletal regions are less accurate due to the absence of critical information in X-ray images such as muscle and fat distribution, which will be discussed in Section \ref{plusandlimi}.

\subsubsection{Quantitative metric-based analysis}
\begin{figure}[htbp]
\centering
\subfloat[]{
\label{PSNRR}
\centering
\includegraphics[width=0.23\textwidth]{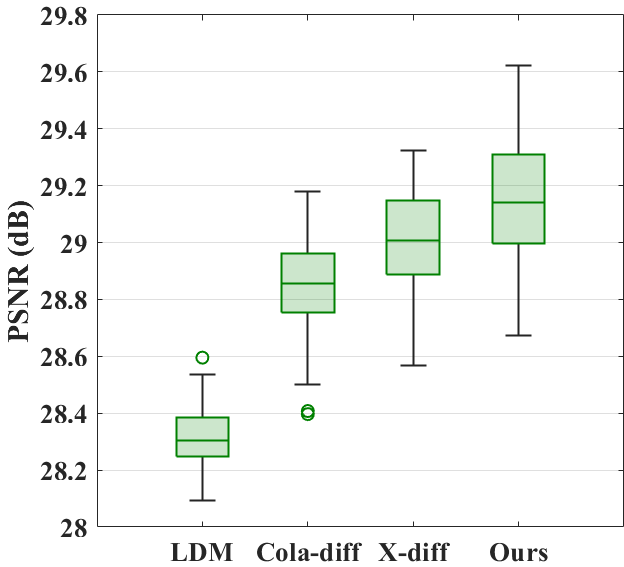}
}
\subfloat[]{
\label{SSIM_RSSIM}
\centering
\includegraphics[width=0.23\textwidth]{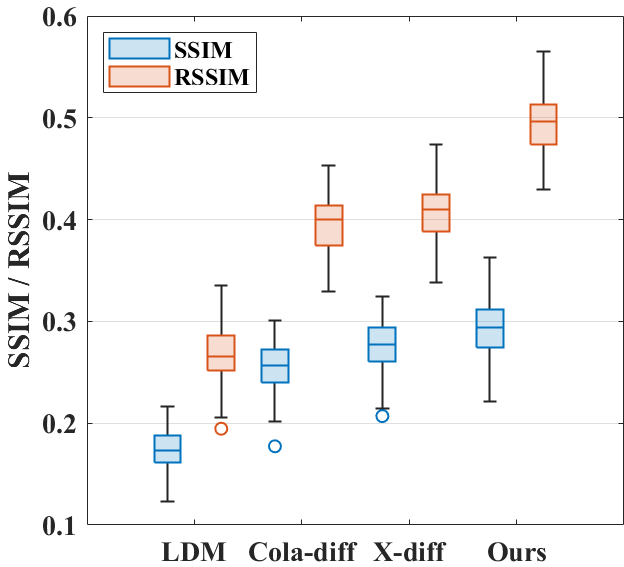}
}
\caption{The box plots visualize the different performance of the evaluated approaches using PSNR \ref{PSNRR} and SSIM/RSSIM \ref{SSIM_RSSIM}.}
\label{boxplots}
\end{figure}

PSNR is a measure used to assess the quality of reconstructed or compressed images compared to their original versions. On the other hand, SSIM is a widely used metric for evaluating the visual similarity between two images considering human visual perception by measuring changes in luminance, contrast, and structural information. However, it does not account for the specific Regions Of Interest (ROI) that are critical in medical imaging. In knee MRI volumes, the most relevant diagnostic information often lies within cartilage, joints, and other areas affected by disease. Therefore, in addition to SSIM, we also adopt a specially designed indicator, the so-called Region-specific Structural Similarity Index (RSSIM), to represent the clinical evaluation of bone morphology. To do so, we utilize the Canny operator with a sigma of 3 to filter and extract bone edges, followed by the application of SSIM. Specifically, we focus on the lower third of the femoral edge to the upper third of the tibial edge to capture ROIs most likely to contain areas of potential pathologies, such as JSN and osteophytes. We compared the performance of the proposed approach with the previously selected methods. Fig. \ref{boxplots} illustrates the statistical distribution of these metrics. As can be seen in Fig. \ref{PSNRR}, our proposed method has the highest median PSNR, indicating better average image quality, with a median slightly above 29.2 dB. Cola-diff and X-diff also demonstrate a high median around 28.9 dB, with LDM being the lowest, around 28.6 dB. This suggests that, based on the PSNR metric, the image quality performance differences among the four methods are relatively minor. On the other hand, Fig. \ref{SSIM_RSSIM} reveals that while our method achieves the highest median SSIM value, approximately 0.28, the differences between our method, Cola-diff, and X-diff are not substantial, mirroring the trend seen with PSNR. However, our method exhibits a significant advantage under the RSSIM metric. With a median RSSIM value close to 0.5, our approach outperforms the closest competitor, X-diff, which only reaches around 0.42, representing a roughly 20$\%$ improvement, which underscores that our approach’s ability to maintain image quality while providing a stronger focus on clinically relevant regions in MRI sequences. 

\subsection{Ablation study on the feature information}
To assess the contribution of the integrated patient-specific radiographic feature information $r$ within the guidance module, we conducted a detailed ablation study, analyzing the impact of specific variables within $r$. This ablation study was performed using the same dataset configuration to ensure consistency and comparability across experiments. Here, we particularly focused on the roles of the variables of chondrocalcinosis (CPPD), attrition, and cysts in Table \ref{features}, given their significant clinical relevance and the challenges associated with interpreting these features from X-ray images alone. For instance, CPPD is associated with the deposition of calcium pyrophosphate crystals in the joint tissues, which manifests as subtle calcifications, which often overlap with normal joint structures, leading to ambiguity in detection using Xrays. Attrition, on the other hand, reflects the loss of articular cartilage and underlying bone, a hallmark of joint degeneration. Its appearance on X-rays can be subtle. Similarly, cystic changes represent fluid-filled cavities or bone erosions that occur in advanced joint diseases, these changes can appear as radiolucent areas on X-rays, and their true extent and clinical implications are difficult to discern without complementary information. By isolating these variables, we aimed to determine the extent to which they enhance the model's ability to synthesize MRI outputs that accurately reflect the underlying pathology. 

As shown in Table \ref{ablation_study}, before integrating the CPPD variable (V00XRCHL), the model generated MRI sequences that entirely lacked any representation of CPPD, as it was unable to incorporate this feature without explicit guidance. However, after integrating CPPD as a guidance variable, the model began to display CPPD in a roughly accurate region (i.e., the lateral compartment). For the attrition, integrating the corresponding variable (V00XRATTL) enabled the model to better capture signs of cartilage loss and bone alterations in the tibia lateral compartment. Similarly, for the cysts, the variable (V00XRCYTL) allowed the model to successfully synthesize fluid-filled regions in the tibia lateral compartment. It is noteworthy that, while the model struggled to fully capture the precise severity and spatial details of certain features, the ablation studies demonstrate that integrating patient-specific radiographic information beyond what X-rays alone provide improves the anatomical realism and clinical relevance of the generated MRI sequences, allows the model to generate MRI outputs that are, to some extent, closer to real MRI scans, which underscores the immense potential of incorporating additional information (e.g.,  Body Mass Index (BMI), neural distribution, etc) as guidance into such models.

\begin{table}[htbp]
\centering
\caption{Ablation studies for feature information}
\label{ablation_study}
\begin{threeparttable}
\setlength{\tabcolsep}{0.2mm}
\begin{tabular}{ccc}
\toprule
Ground Truth &  with CPPD$^1$  & without CPPD \\
\midrule
\begin{minipage}[b]{0.325\columnwidth}\centering \raisebox{-.5\height}{\includegraphics[width=\linewidth]{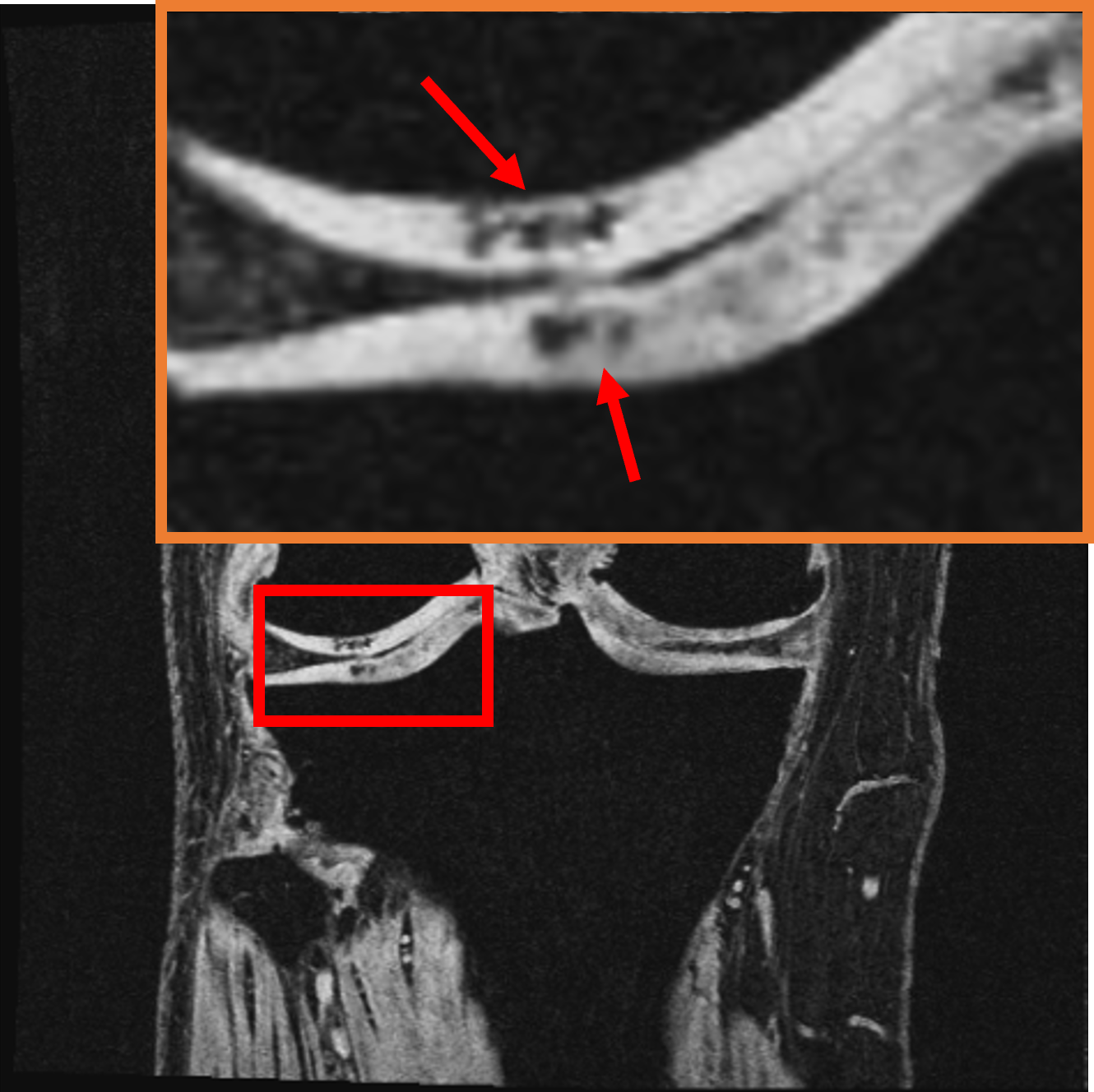}}\end{minipage}&\begin{minipage}[b]{0.325\columnwidth}\centering \raisebox{-.5\height}{\includegraphics[width=\linewidth]{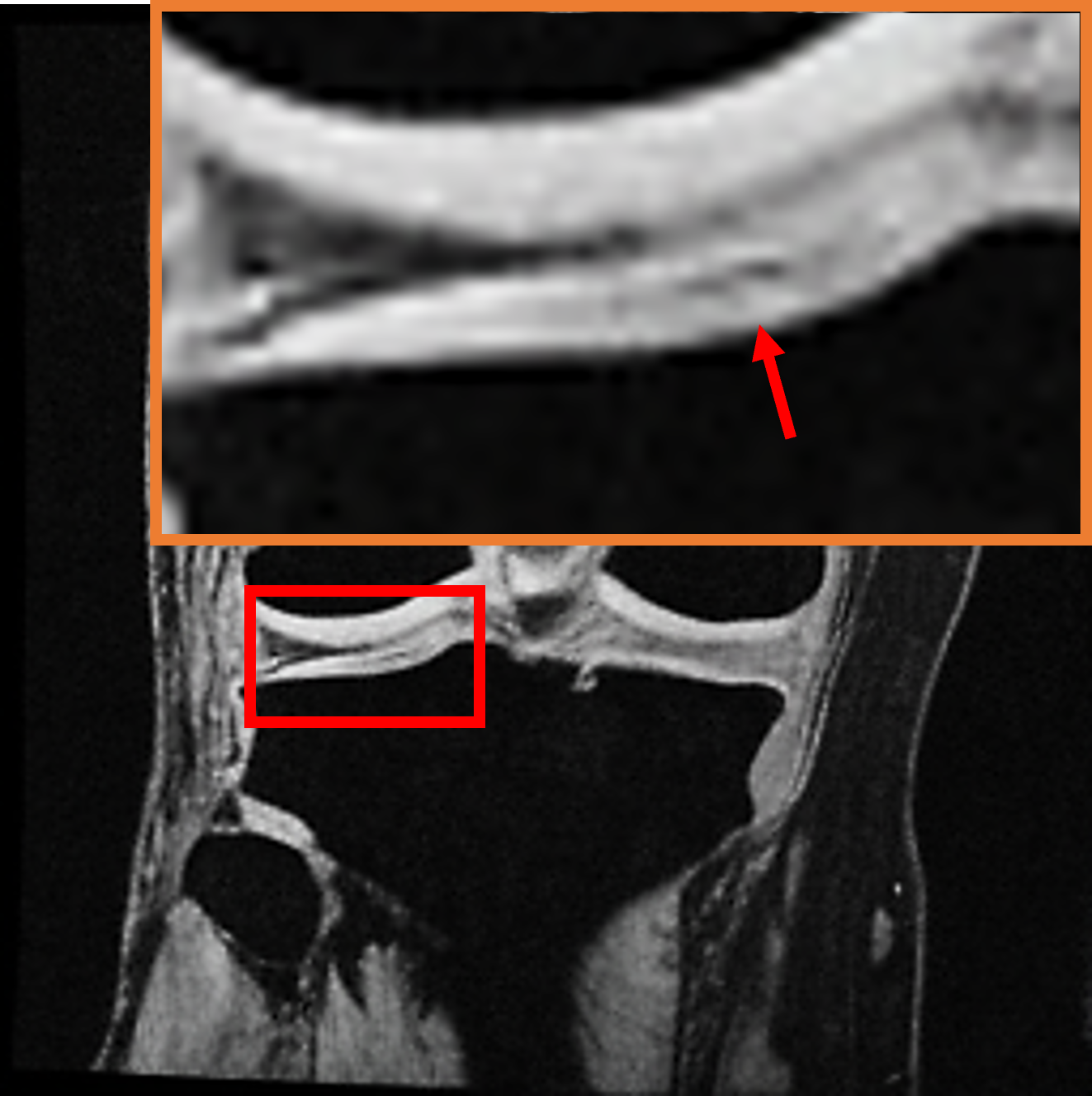}}\end{minipage}&\begin{minipage}[b]{0.325\columnwidth}\centering \raisebox{-.5\height}{\includegraphics[width=\linewidth]{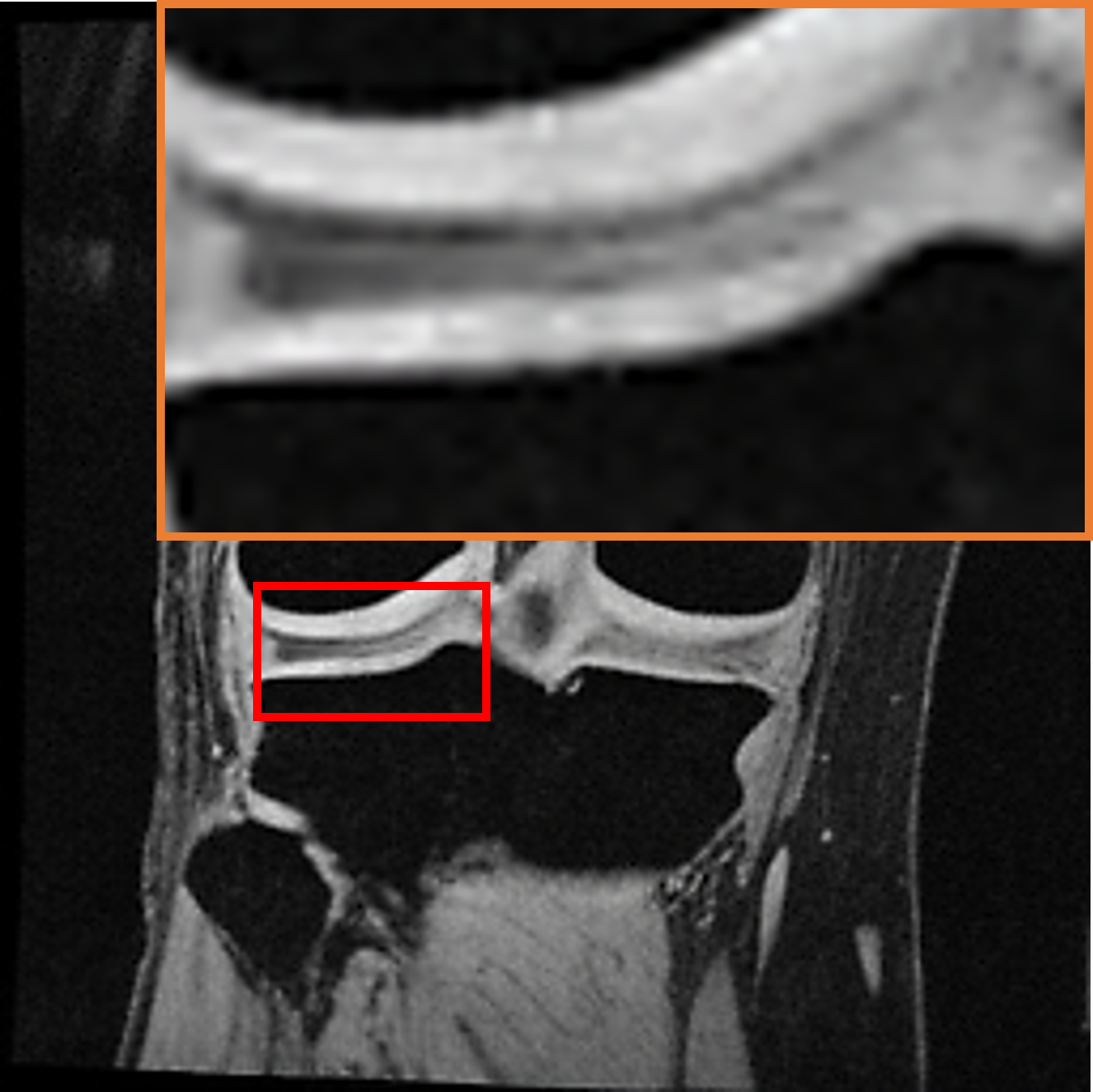}}\end{minipage}\\
\midrule
Ground Truth & with Cysts$^3$ & without Cysts\\
\midrule
\begin{minipage}[b]{0.325\columnwidth}\centering \raisebox{-.5\height}{\includegraphics[width=\linewidth]{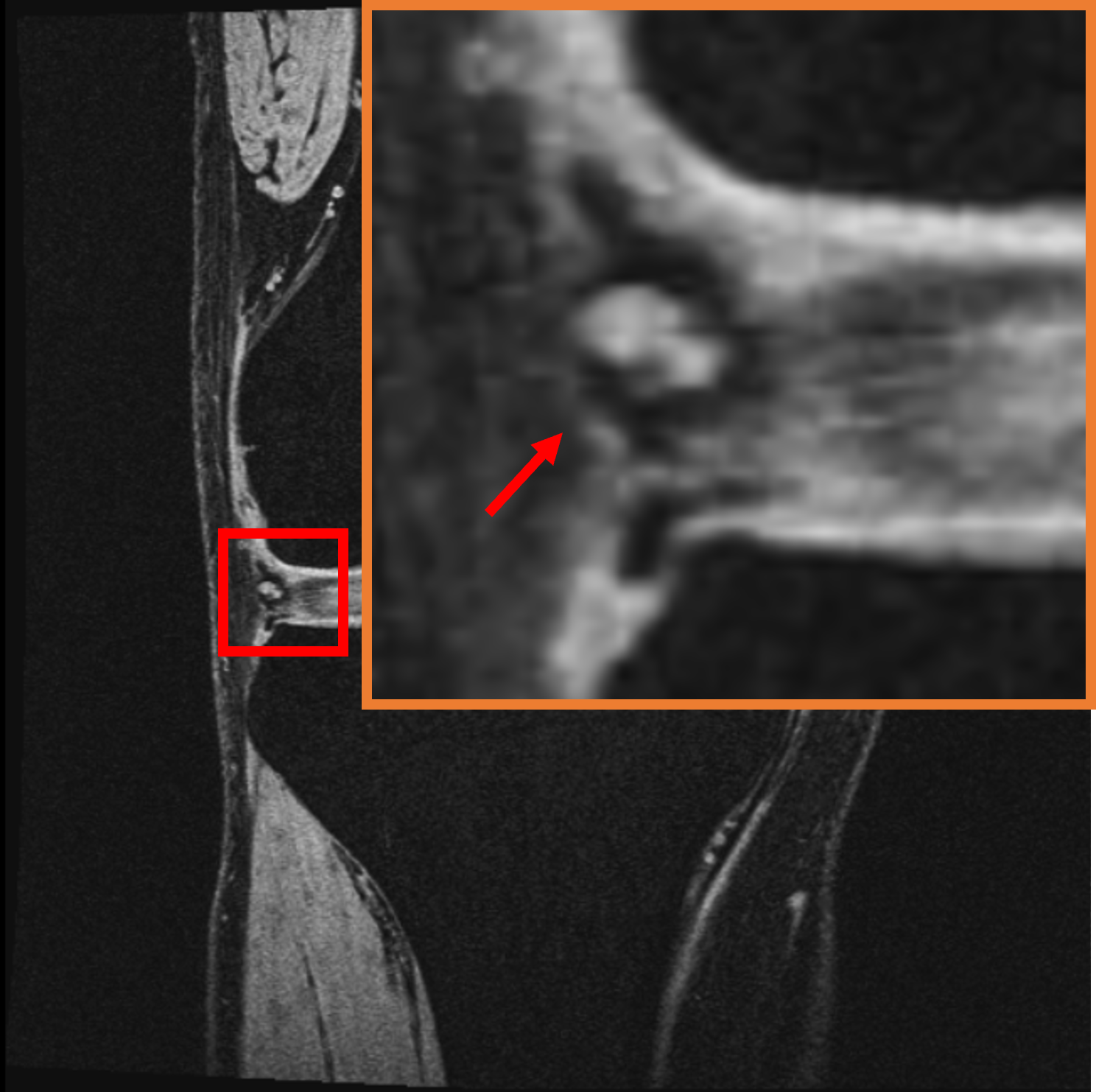}}\end{minipage}&\begin{minipage}[b]{0.325\columnwidth}\centering \raisebox{-.5\height}{\includegraphics[width=\linewidth]{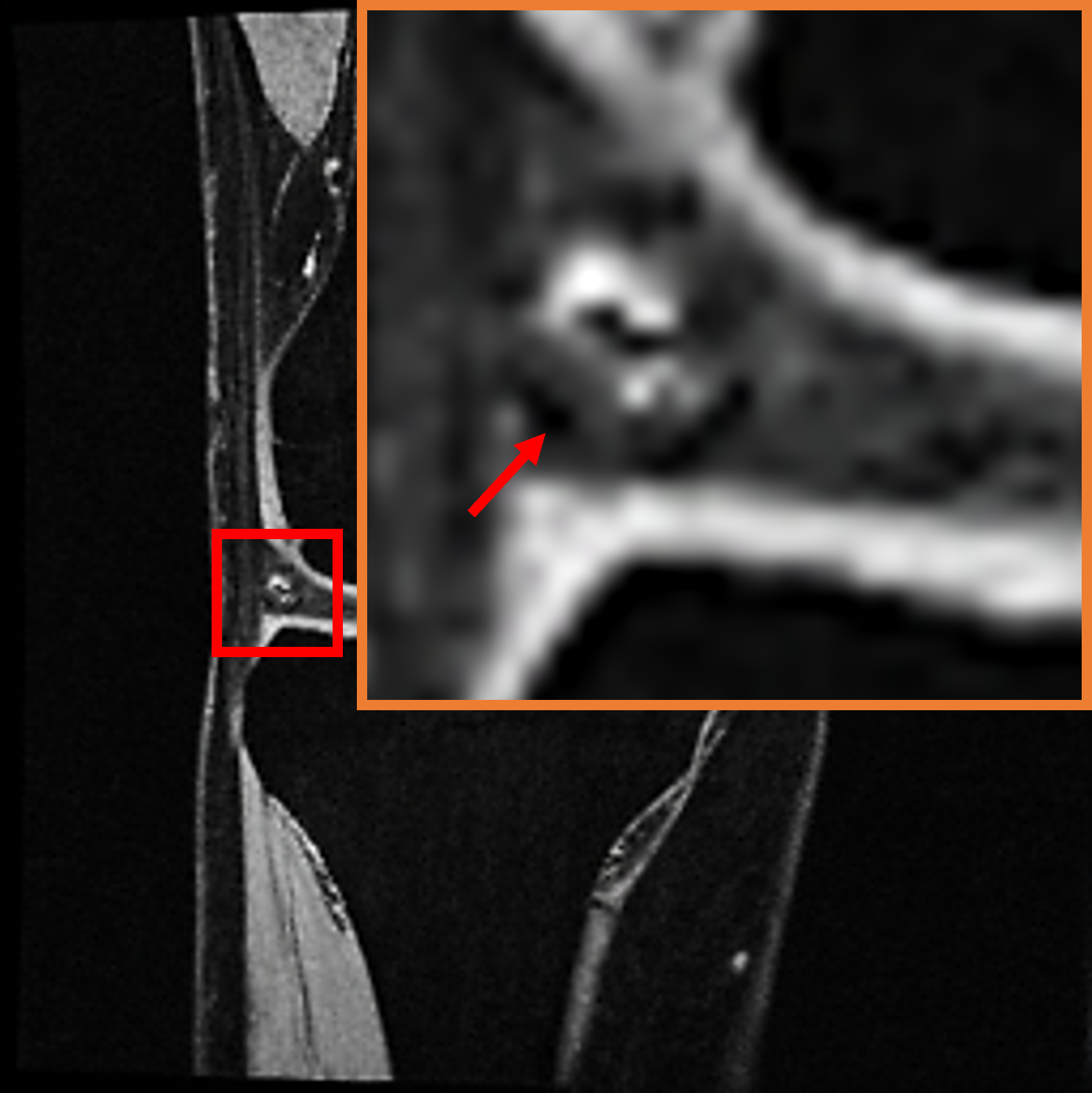}}\end{minipage}&\begin{minipage}[b]{0.325\columnwidth}\centering \raisebox{-.5\height}{\includegraphics[width=\linewidth]{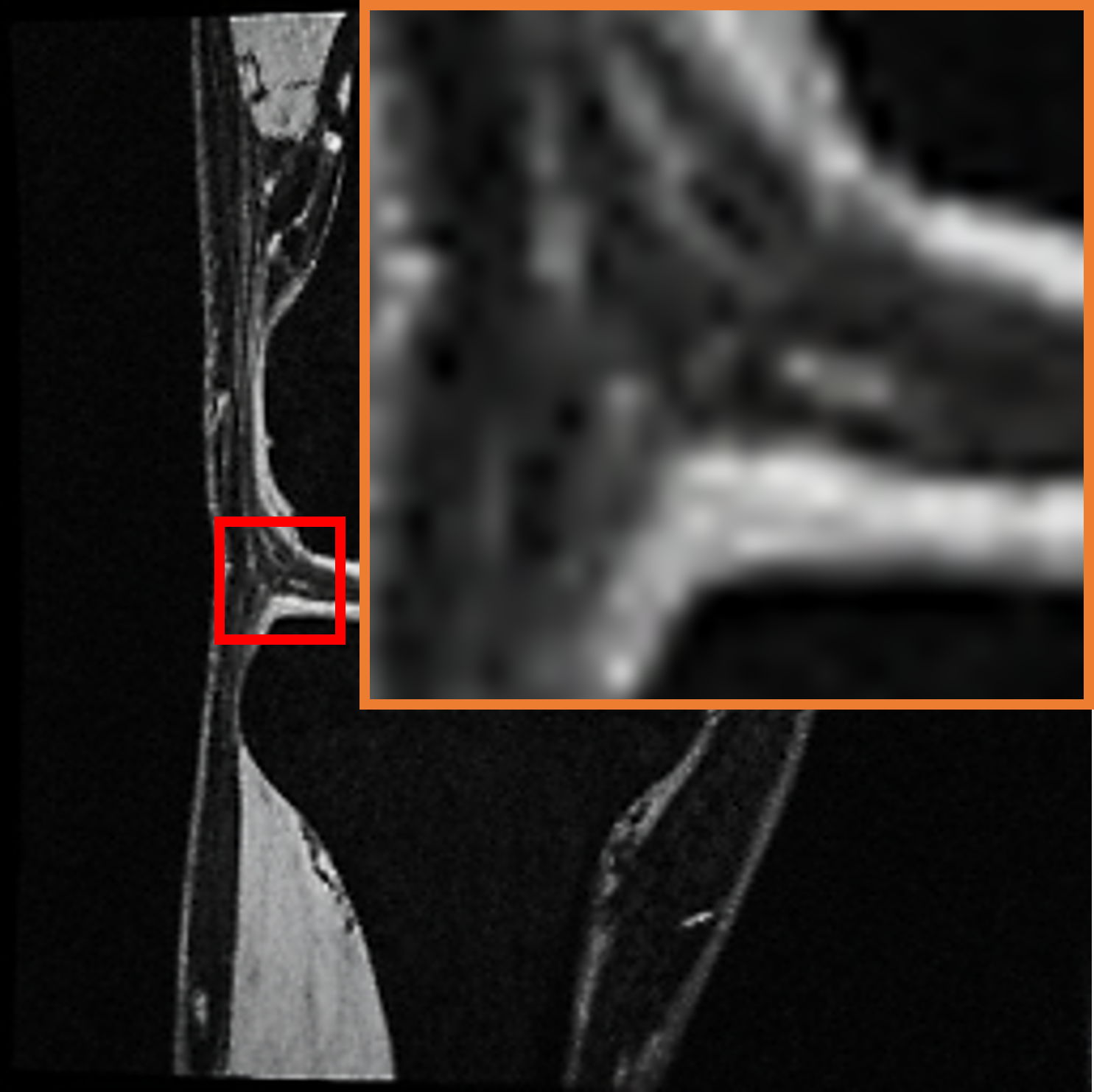}}\end{minipage}\\
\midrule
Ground Truth &  with Attrition$^2$  & without Attrition \\
\midrule
\begin{minipage}[b]{0.325\columnwidth}\centering \raisebox{-.5\height}{\includegraphics[width=\linewidth]{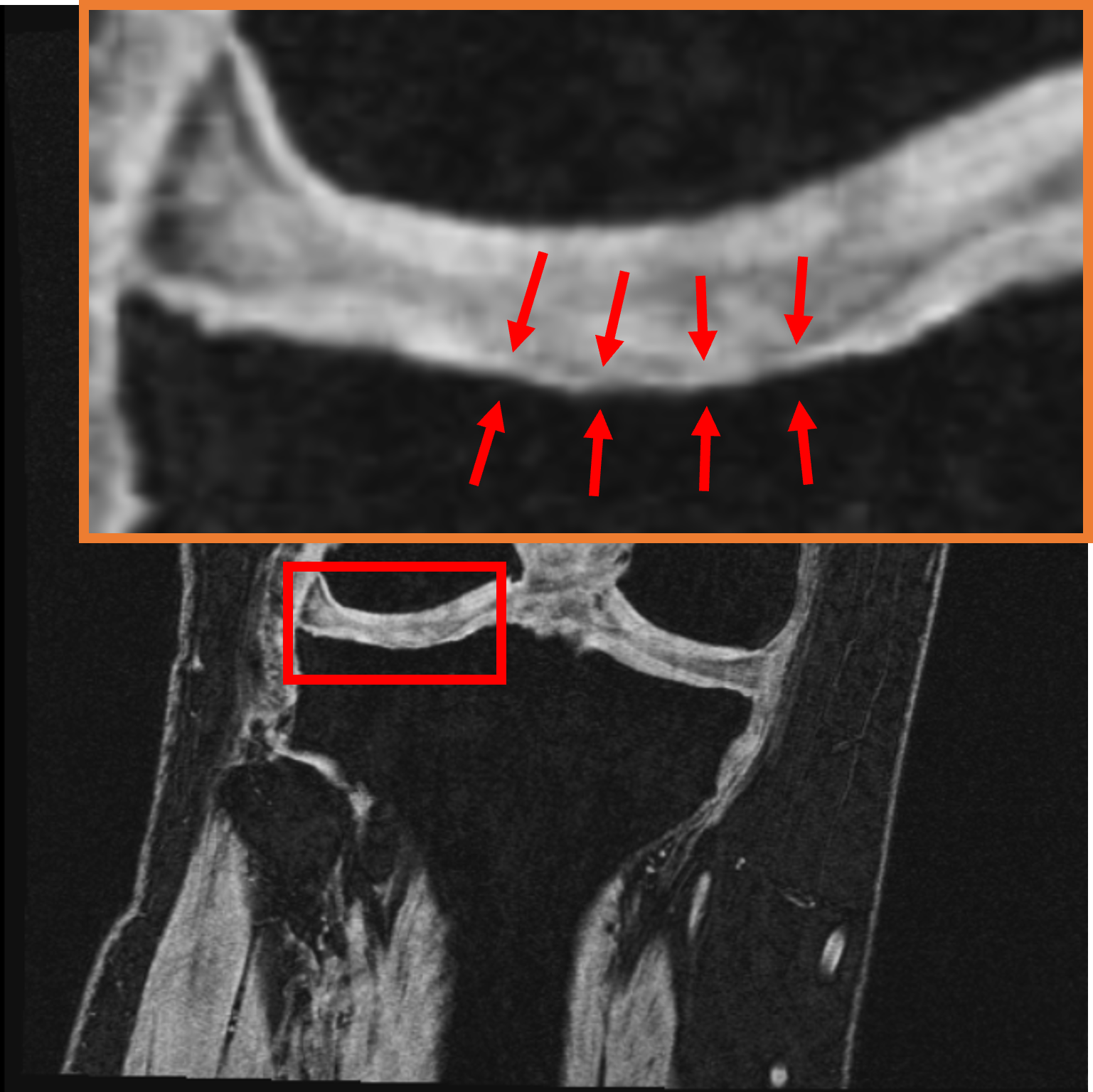}}\end{minipage}&\begin{minipage}[b]{0.325\columnwidth}\centering \raisebox{-.5\height}{\includegraphics[width=\linewidth]{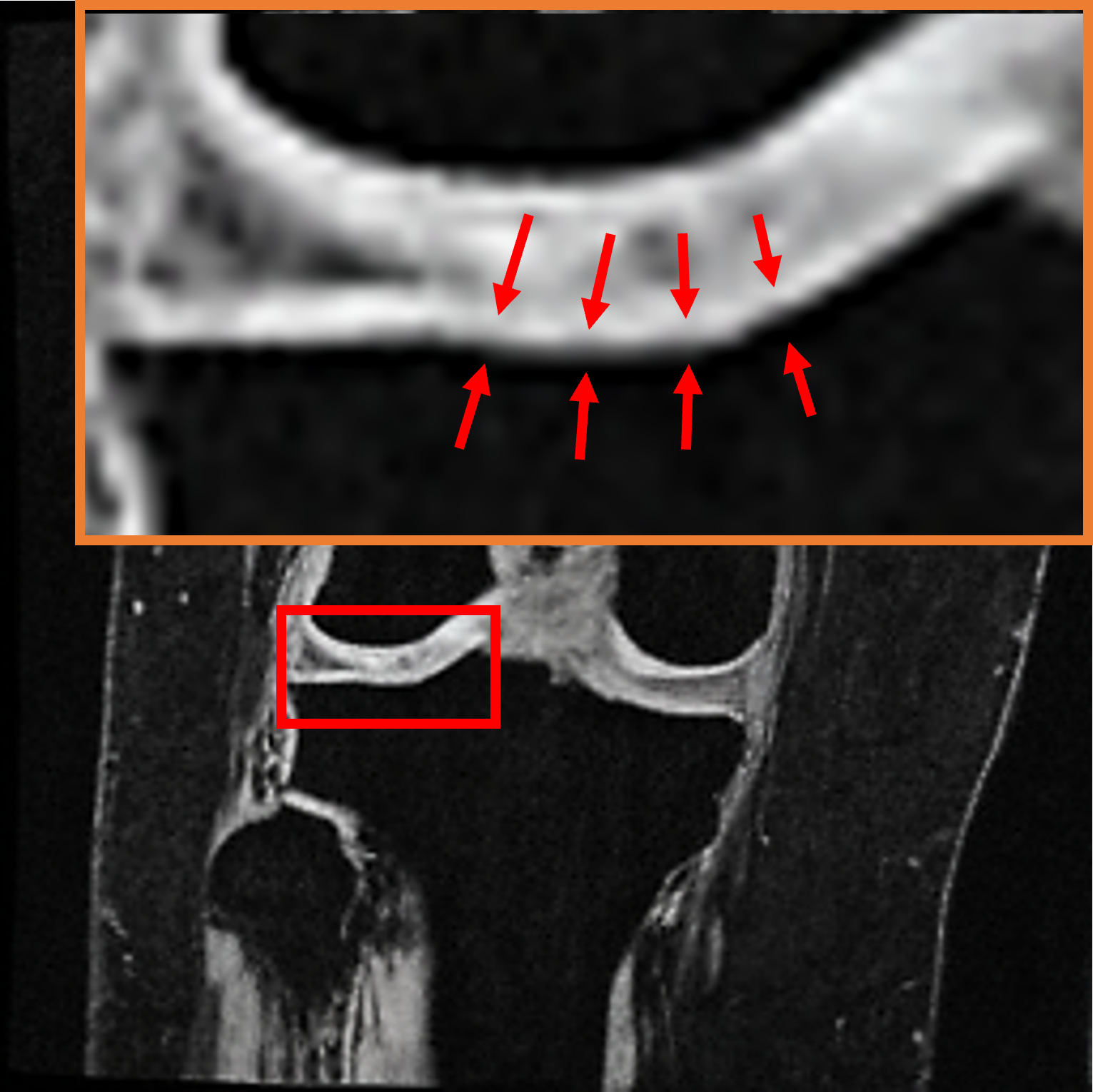}}\end{minipage}&\begin{minipage}[b]{0.325\columnwidth}\centering \raisebox{-.5\height}{\includegraphics[width=\linewidth]{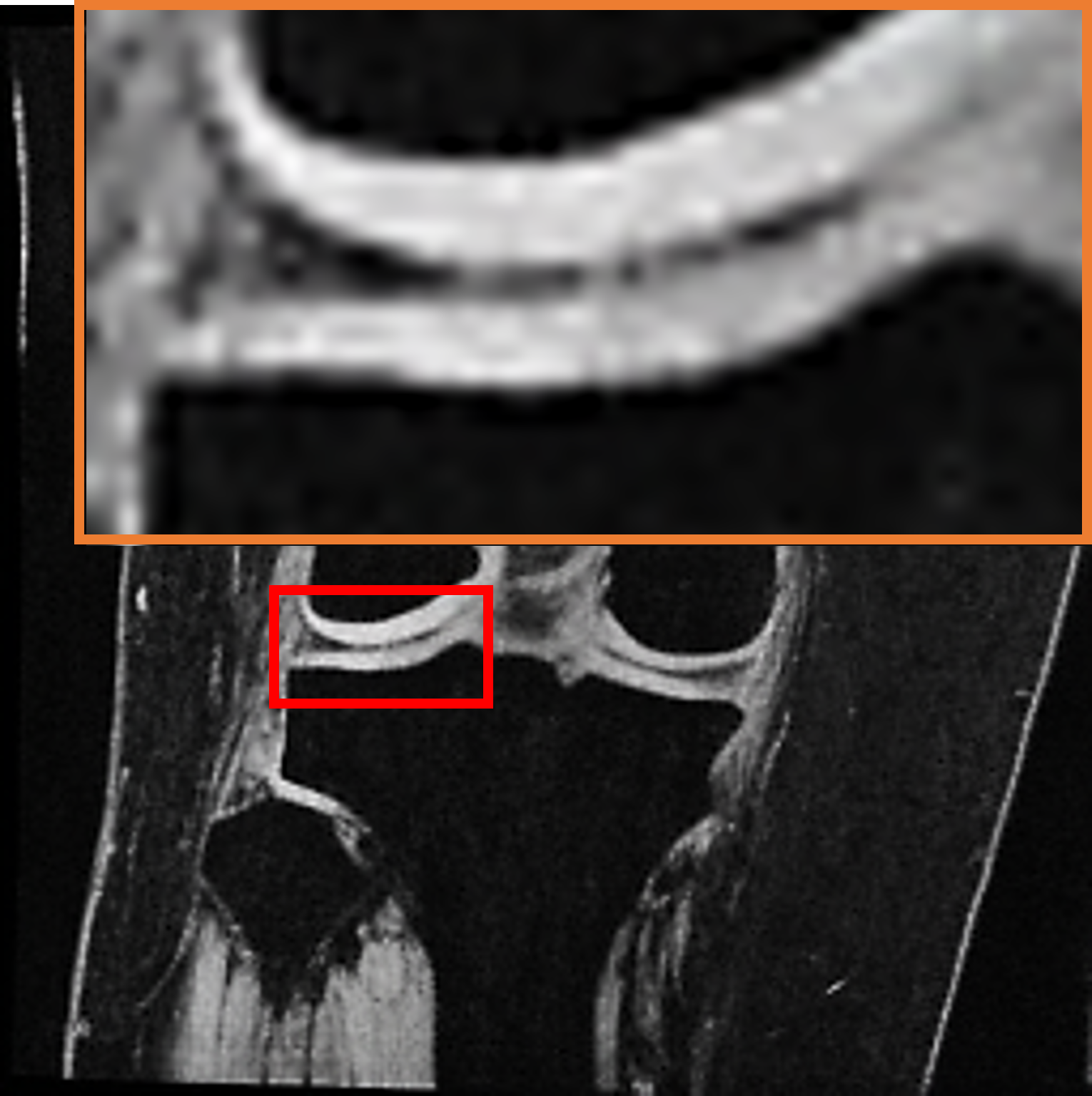}}\end{minipage}\\
\bottomrule
\end{tabular}
\begin{tablenotes}
\footnotesize
\item[$1$] V00XRCHL=1: CPPD present in lateral compartment
\item[$2$] V00XRCYTL=1: Cysts present in the tibia lateral compartment
\item[$3$] V00XRATTL=1: Attrition present in tibia lateral compartment
\end{tablenotes}
\end{threeparttable}
\end{table}

\begin{table*}
\centering
\caption{Visualization of the synthesized intermediate frames}
\label{vlisualization_500}
\setlength{\tabcolsep}{0.52mm}
\begin{tabular}{cccccc}
\toprule
Starting image (real) & \multicolumn{4}{c}{$\stackrel{\text{\footnotesize Synthesized intermediate frames}}{\makebox[10cm]{\rightarrowfill}}$} & Ending image (real)\\
\midrule
\begin{minipage}[b]{0.31\columnwidth}\centering \raisebox{-.5\height}{\includegraphics[width=\linewidth]{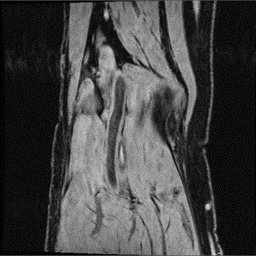}}\end{minipage}&\begin{minipage}[b]{0.31\columnwidth}\centering \raisebox{-.5\height}{\includegraphics[width=\linewidth]{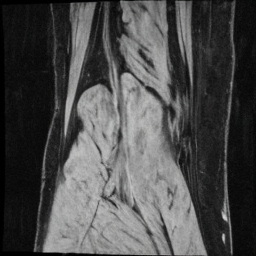}}\end{minipage}&\begin{minipage}[b]{0.31\columnwidth}\centering \raisebox{-.5\height}{\includegraphics[width=\linewidth]{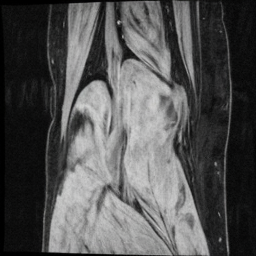}}\end{minipage}&\begin{minipage}[b]{0.31\columnwidth}\centering \raisebox{-.5\height}{\includegraphics[width=\linewidth]{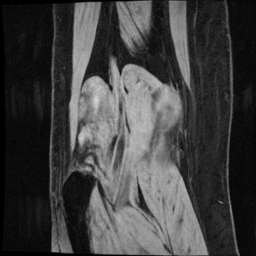}}\end{minipage}&\begin{minipage}[b]{0.31\columnwidth}\centering \raisebox{-.5\height}{\includegraphics[width=\linewidth]{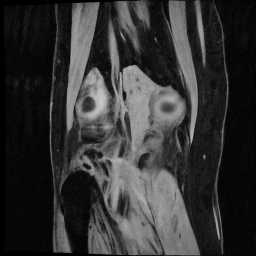}}\end{minipage}&\begin{minipage}[b]{0.31\columnwidth}\centering \raisebox{-.5\height}{\includegraphics[width=\linewidth]{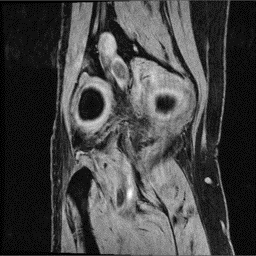}}\end{minipage}\\
\bottomrule
\end{tabular}
\end{table*}

\subsection{Impact of the number of inference steps}
\begin{figure}[htbp]
\centering
\includegraphics[width=0.385\textwidth]{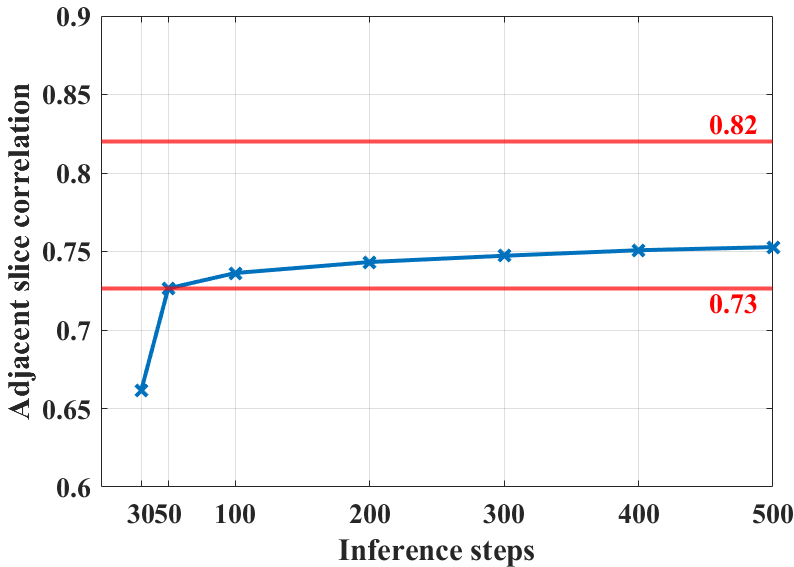}
\caption{The adjacent slice correlation coefficients under different inference steps.}
\label{corr}
\end{figure}

For the inference steps $s$, in addition to adhering to the original depth setting (i.e., $s=S=50$), we also experimented with varying inference steps (i.e., 30, 100, 200, 300, 400, and 500), to explore the impact of increasing the number of inference steps on the quality and accuracy of the generated slices. Fig. \ref{corr} shows the adjacent slice correlation coefficients under different inference steps. As can be seen, the red horizontal lines at 0.73 and 0.82 represent the mean adjacent slice correlation coefficients of the generated MRI matching the volume of the original MRI (i.e., $s=S=50$) and that of the original MRI itself, respectively, with the latter serving as the upper benchmark for comparison. At 30 inference steps, the adjacent slice correlation starts at approximately 0.66, reflecting the impact of downsampling, which introduces inconsistencies and reduces the alignment between neighbouring slices. As the number of inference steps increases to 50, where the depth aligns with that of the original MRI volume, the correlation coefficient jumps sharply to around 0.73, suggesting a significant improvement in slice coherence. At 100 steps, the correlation continues to increase modestly. Beyond 100 steps, the improvement becomes more gradual as the number of inference steps increases, with diminishing returns, eventually stabilizing around 0.76 by 500 inference steps. Table \ref{vlisualization_500} illustrates the synthesized intermediate frames between the original adjacent images (i.e., starting and ending images) using 500 inference steps. A distinct representative stage is evaluated: The transition from predominantly muscle tissue to the initial appearance of bone. In this stage, the synthesized intermediate frames display relatively smooth and continuous transitions between bone and soft tissue structures, which underscores the robustness and effectiveness of these interpolation frames.

The above results demonstrate that increasing the number of inference steps $s$ enhances interpolation, showcasing its potential for improving temporal resolution. By generating additional intermediate frames from a smaller set of real frames, interpolation effectively offers more continuous data without lengthening the actual scan duration. On the other hand,  despite the improvements observed, the correlation coefficient from the interpolated MRI fails to reach the 0.82 benchmark of the original MRI volume, even with the maximum 500 inference steps, which indicates that, while the interpolation process can enhance the consistency between adjacent slices, inherent limitations remain. These limitations arise from the inherent constraints of using X-rays as the sole imaging source of information, which will be discussed in Section \ref{plusandlimi}.

\subsection{Discussion}
In this paper, we introduce a novel diffusion-based model that generates full MRI volumes from a single X-ray image. Specifically, the model leverages conditional inputs from X-rays, combined with target depth and patient-specific feature information, to guide the synthesis process for the MRI slice generation. Experimental results demonstrate that, compared to SOTA methods, our approach generates pseudo-MRI sequences that are closer to real MRI scans in terms of image quality and clinical relevance. Furthermore, ablation studies on the guidance module validate that incorporating supplementary information beyond what is available from X-rays alone improves the accuracy of the generated MRI, which highlights the immense potential for integrating additional independent information in the future to further enhance MRI generation performance. Lastly, by increasing the number of inference steps, the model exhibits effective interpolation, resulting in smoother transitions between slices. Several points are to be discussed.
\subsubsection{How it works?}
In this approach, we move away from focusing on the distinct imaging principles and techniques of X-ray and MRI. Instead, we only concentrate on the mapping relationship between these two imaging modalities. Much like an extremely experienced senior radiologist who, when viewing an X-ray image, could to some extent mentally infer what the corresponding MRI scans might reveal, which is our proposed AI model's aim. To this end, for the model's training, vast amounts of high-quality paired X-ray and MRI have significantly propelled this.

\subsubsection{Why use a 2D diffusion model instead of a 3D one directly?}
We opted to use a 2D diffusion model, generating individual MRI slices at different target depths during the inference phase, which are then stacked to form a 3D MRI sequence. By using this method, the model can effectively learn and capture the spatial relationships between the individual slices, allowing it to understand inter-slice dependencies. Additionally, employing a 2D model reduces computational complexity, making both the training and inference processes more efficient compared to handling full 3D data. This approach also enhances the interpretability of the entire process, as it enables a more granular analysis of how each slice contributes to the final 3D reconstruction.

\subsubsection{Strengths and limitations}
\label{plusandlimi}
This study has several notable strengths. Firstly, it introduces an innovative method for generating MRI from one single X-ray image, attempting to bridge the gap between these two imaging modalities using advanced AI technology, which represents a novel attempt in cases where MRI may not be readily available. Our experiments demonstrate that incorporating supplementary information (i.e., patient-specific radiographic feature information in this study) beyond the X-ray images themselves enhances the quality of the generated MRI, which underscores the potential of integrating more diverse information sources in the future to produce more accurate generated MRI volumes. Second, the study employs a 2D diffusion model, which not only reduces the computational resources required but also allows for more detailed depth control over the generated images, leading to a more interpretable and transparent reconstruction process. Finally, the study’s focus on clinically relevant metrics, including both traditional image quality assessments and pathology-specific evaluations, ensures that the generated MRI are not only visually similar to ground truth but also clinically relevant. This study has several limitations that warrant consideration. Our model are highly dependent on the availability of large, high-quality paired datasets. The scarcity of such datasets, particularly in certain clinical scenarios or for rare diseases, may limit the generalizability of our approach. Moreover, the computational resources required for training, despite being reduced by using a 2D model, high-performance hardware is still necessary, which presents a challenge for broader implementation in clinical settings. Therefore, more additional information (e.g., Body Mass Index (BMI), age, gender, etc) could be integrated to further enhance the generation performance. To improve the model’s deployability, the use of knowledge distillation could be of interest.

\section{Statements}
This manuscript was prepared using data from the OAI. The views expressed in it are those of the authors and do not necessarily reflect the opinions of the OAI investigators, the National Institutes of Health (NIH), or the private funding partners.

\section{Acknowledgements}
The authors gratefully acknowledge the support of the Ralph Schlaeger Research Fellowship under award number 246448 from Massachusetts General Hospital (MGH), Harvard Medical School (HMS) and the French National Agency of Research (ANR) under project number ANR-20-CE45-0013-01.

\bibliographystyle{IEEEtran}
\bibliography{references}

\end{document}